\documentclass[12pt]{iopart_comp_amsmath}

\usepackage{amsmath}
\usepackage{amssymb}

\usepackage{graphicx}

\usepackage{tensor}
\newcommand{\te}[2]{\tensor{#1}{#2}}

\newcommand \be{\begin{eqnarray}}
\newcommand \ee{\end{eqnarray}}
\newcommand \ba{\begin{align}}

\newcommand {\p}[1]{\partial_{#1}}

\newcommand{\ch}[2]{\genfrac{\{}{\}}{0pt}{}{#1}{#2}}

\begin{document}

\title{Consistent solution of Einstein-Cartan equations with torsion outside matter}

\author{Klaus Morawetz}

\address{M\"unster University of Applied Sciences,
Stegerwaldstrasse 39, 48565 Steinfurt, Germany\\
International Institute of Physics- UFRN,
Campus Universit\'ario Lagoa nova,
59078-970 Natal, Brazil}
\ead{morawetz@fh-muenster.de}
\vspace{10pt}
\begin{indented}
\item[]November 2026
\end{indented}

\begin{abstract}
The Einstein-Cartan equations in first-order action of torsion are considered. From Belinfante-Rosenfeld equation special consistence conditions are derived for the torsion parameters relating them to the metric. Inside matter the torsion is given by the spin which leads to an extended Oppenhaimer-Volkov equation. Outside matter a second solution is found besides the torsion-free Schwarzschild one with the torsion completely determined by the metric and vice-versa. This solution is shown to be of non-spherical origin and its uniqueness with respect to the consistence is demonstrated. Unusual properties are discussed in different coordinate systems where the cosmological constant assumes the role of the Friedman parameter in Friedman-Lama\^itre-Robertson-Walker cosmoses. Parameters are specified where wormholes are possible. Transformations are presented to explore and map regions of expanding and contracting universes to the form of static metrics. The autoparallel equations are solved exactly and compared with geodesic motion. The Weyl tensor reveals that the here found solution is of Petrov-D type.   \end{abstract}

%
%
\submitto{corrected version of \CQG 36 (2021) 205003}
%
\maketitle

%
%

\section{Introduction}

Exact solutions of Einstein equations have a long history, to mention only the most known ones like the  inner and outer Schwarzschild solution \cite{Sch16,Sch16a}, the Friedman-Lama\^itre-Robertson-Walker (FLRW) metric \cite{F22,F24,La31,Ro36,Wa37} as well as Goedel's universe \cite{Go47}. There are many reviews and text books classifying these unfoldings in a systematic way \cite{Ah19,StK03,GP09}. Especially the formulation in tetrads formalism \cite{Pi57,NP62,MTW73} has paved the way to find new exact results.

These solutions are forming the standard cosmological model considering the curvature but neglecting any torsion. This has been included to modify the Einstein gravitation theory by Cartan \cite{Cart23} which together with the gauge-invariance principle \cite{Wey29,SC64} have yield a practicable classical field theory.
One can show that as localization of the Poincar\'e group, the Riemann-Cartan
and Weyl-Cartan differential geometry follows \cite{BFZ06}.
 For an overview about the development of these extensions see \cite{Hehl76,Ham02} and for a modern mathematical presentation \cite{shap02,Pop09}. The inclusion of torsion has regained a current interest since it promises a new direction for the search beyond the standard cosmological model \cite{HHKD74}. It has been even considered as a candidate for dark matter \cite{Pop11,BMS17,Mg19} and was suggested to explain matter-antimatter imbalance \cite{Pop11}. 

The Einstein-Cartan cosmologies \cite{MNB19} show that no singularity appears \cite{Kuch78,Pop12} replacing the Big Bang by a big bounce \cite{Br08,Ber09,Pop12,pop18,UP19}. As a consequence a black-hole cosmology results \cite{Pop16} which could unify the big bounce and inflation \cite{Gas86,pop18,Cu20}. Here the torsion generates a gravitational repulsion which avoids initial singularities \cite{POP10}. The emergent scenario has been investigated to find a stable solution of Einstein-Cartan equations only for the closed universe \cite{HWH15}. Avoiding such initial singularities allows also to study primordial fluctuations and a finite period of cosmic inflation can be rooted to particle production due to curved space time \cite{pop18a}. The creation of magnetic fields during that inflation period requires a coupling between torsion and electromagnetism \cite{KSP20}.  For late cosmology a non-adiabatic expansion was found and that the second law of thermodynamics requires a positive torsion term \cite{CIL20}. Asymptotic flat solutions have been considered with respect to gravitational lensing in \cite{CZJ18} and rotating and expanding solutions are presented in \cite{Gal09}. The inner star objects with spin and torsion are considered in \cite{LS19} and mass bounds in \cite{BHT16}.

The origin of torsion is considered to be the spin \cite{Ham99}. This has been pronounced recently \cite{Hamm18} by showing that covariant electromagnetic gauge-invariance leads to torsion. Furthermore, the Einstein-Cartan-Kibble-Sciama theory \cite{Hehl76} allows to formulate the Einstein-Maxwell-Dirac theory in geometric form \cite{pop10a}. Gauge covariant generalizations are discussed in \cite{DY19} and an extended conformal invariance and non-metricity are found in \cite{KDD12,KL20}. Nonsymmetric metric tensor has been used then to explore gravity and spin \cite{Hamm19}. The metric and connection as fundamental quantities to form the spacetime have been constructed in a Weyl-invariant way \cite{VTM15}. A recent classification of metric-affine theories can be found in \cite{IK19}. The Einstein–Hilbert
theory itself was obtained as an effective theory due to quantum corrections of torsion
with the conjecture that torsion is of intrinsic quantum nature \cite{KP08}.

The possibility to observe torsion effects at CERN LHC has been discussed in \cite{BS99,BSV07}. The perturbative effects of torsion to the secular perihelion precession has been predicted to be observable for Mercury \cite{Ac15}. Experimental bounds to observe space-time torsion are suggested by spin-torsion coupling \cite{OST14}. Cosmological signatures of torsion are discussed recently \cite{KRB20}. It is shown that the angular diameter and luminosity distance become different due to torsion and it was predicted that torsion should be visible in a lower redshift which might change the interpretation of supernovae data.

The exact solutions of Einstein-Cartan theory are based on different Lagrangians considering the contortion in different forms. Mostly the Weyssenhoff spin-fluid assumption \cite{We47} about the structure of torsion is used \cite{RS83,Ob87,Br07,Ber09,Ka10,ZMRS14,MSS16,SZH18}. An averaging about randomly oriented spins leads basically to a shift in momentum and energy densities \cite{Pop16}.  A string approach resulting into slightly different equations of motions of Einstein-Cartan-Kalb-Raimond coupling can be found in \cite{FGLMY12}. The tetrad formalism allows to construct systematically solutions \cite{Ka10,MSS16,Ka19} not restricted to spin-liquids. The Einstein-Cartan-Kibble-Sciama scheme was solved in \cite{SZH18} with higher-order terms of contortion tensor. Such higher-order terms like quartic terms appear also in bigravity approaches \cite{DN19}. Demanding the sum of the curvature and torsion scalar to vanish in a background, Weyl-Cartan space-time has been considered as teleparallel gravity models \cite{HHSS13}. Special attention has been also paid to time-evolving wormhole solutions \cite{MZH17,MZH19}. 

All these solutions have in common that the torsion is considered as created by matter spin and consequently vanishes if the spin is absent. It is sometimes reasoned that the torsion field does not propagate itself and therefore it is a byproduct of curvature created by the energy-momentum tensor. This is underlined by the observation that static spherically symmetric solutions are not possible for propagating torsion \cite{Hamm91}. Consequently, the dynamics of metric-affine gravity with torsion is discussed in \cite{VIT11}. However, there are reports about the metric of the axially symmetric solution \cite{BKF16,MKK20} in empty space and its application to the rotation curves of spiral galaxies \cite{BKF16}. It was found that the vacuum torsion interacts with spin momenta of astrophysical objects which can lead to modifications of Newton's law \cite{MKK20}.

We will show in this paper that a non-propagating static solution with torsion exists outside of matter. It will be found that the torsion itself is appearing from Einstein-Cartan equations even without matter and the presence of spin. The mere fact of allowing non-propagating torsion leads to equations which show besides the Schwarzschild solution of absent torsion a second one. This second solution does not match with the Schwarzschild solution since the torsion parameter itself becomes fixed by the curvature and vice-versa. It will be shown that this presented solution is distinct from the known ones and possesses unusual properties.

The outline of the paper is as follows. In the next chapter the main formulas of Einstein Cartan theory are shortly summarized. In the third chapter we explore the Belinfante-Rosenfeld equation \cite{BEL40,Ros40} which will provide consistence constraints on possible torsion terms mostly not considered. The Weyssenhoff fluid assumption without averaging about torsion variables will be used further. In chapter IV we present the internal and external Schwarzschild solution with the main result of a second external solution disconnected from the Schwarzschild one. A detailed comparison with other known solutions is performed in order to show their difference. In chapter V. we extend this solution to possible non-spherical ones and proof the uniqueness of this static solution. The properties of the found solution is explored in chapter VI where it is transformed into different coordinate systems to reveal different properties. The possible wormholes and regions of FLRW metrics are discussed there. The autoparallel equations are solved exactly and compared with geodesic motion.  At the end of this chapter finally the Weyl tensor is calculated and the solution is identified as Petrov-D type. Chapter VII summarizes and concludes.

\section{General formulas}

We work in the metric notation $diag(1,-1,-1,-1)$. The
Einstein-Hilbert action with the extension to torsion 
\cite{Cart23,Wey29,Hehl76,Pop09}
\be
{\cal L}&=&-{1\over 2\kappa}\int \left [P-g^{ik}\left (\te{C}{^j_i_j}\te{C}{^l_k_l}-\te{C}{^l_i_m}\te{C}{^m_k_l}\right )\right ]\sqrt{|g|}d\Omega
\nonumber\\
&&+{\cal L}_m
\label{1}
\ee
is given in terms of the Riemann curvature tensor $P$, the metric $g$ and the contortion tensor $C$. The latter one gives the relation between affine connection $\Gamma$ and the Levi-Civita connection or Christoffel symbols $\{\}$ as
\be
\te{\Gamma}{_i_j^k}= \ch{k}{ij}+\te{C}{^k_i_j}
\label{GC}
\ee
with
\be
\ch{k}{ij}=\frac 1 2 g^{km}\left (
\p i g_{jk}+ \p j g_{im}-\p m g_{ij}
\right )
\ee
due to the metric compatibility, $\nabla_k g_{ij}=0$ which in any subsequent calculation will be ensured. We restrict to lowest order in the action, for higher orders see \cite{VIT11}.

The variation of the matter part of the Lagrangian (\ref{1}) defines the metric dynamical energy-momentum tensor ${\cal T}$ as
\be
\delta {\cal L}_m=-\frac 1 2 \int {\cal T}^{ij}\delta g_{ij} d\Omega.
\ee
The Belinfante Rosenfeld equation \cite{BEL40,Ros40} relates the dynamical metric ${\cal T}$ and the canonical energy-momentum $T$ tensor by the torsion tensor $S$ and the spin tensor $s={2\over \sqrt{|g|}}{\delta \cal L\over \delta C}$
\be
{\cal T}_{ik}&=&T_{ik}+\varepsilon Z_{ik}
\nonumber\\
Z_{ik}&=&-\frac 1 2 (\nabla_l-2 S_l) (\te{s}{_{ik}^l}-\te{s}{_k^l_i}+\te{s}{^l_{ik}})
\label{BR}
\ee
where $S_i=\te{S}{^k_i_k}$ and the contortion tensor is linked to the torsion tensor by
\be
\te{C}{_k_i_j}=\te{S}{_k_i_j}+2 \te{S}{_{(ij)}_k}.
\label{C}
\ee
In order to keep track of the contribution by the Belinfante-Rosenfeld equation $Z_{ik}$, we denote it by an auxiliary factor $\varepsilon=1$. This allows to discuss different cases in the literature, e.g. neglecting it $(\varepsilon=0)$ and tracking the dynamical and canonical energy-momentum tensor.
Here the covariant derivative $\nabla$ with respect to the affine connection is related to the covariant derivative with respect to the Levi-Civita connection $\nabla^{\{\}}$ by
\be
\nabla_k A^l&=&\nabla_k^{\{\}} A^l+\te{C}{^l_m_k} A^m
\nonumber\\
\nabla_k A_l&=&\nabla_k^{\{\}} A_l-\te{C}{^m_l_k} A_m.
\label{rule}
\ee

The Einstein-Cartan equations as variation of (\ref{1}) with respect to the metric tensor $g$ firstly describe the connection of the Riemann tensor to the dynamical metric tensor
\be
G_{ik}=P_{ik}-\left (\lambda+\frac P 2\right ) g_{ik}=\kappa {T}_{ik}+\kappa \varepsilon Z_{ik}+\kappa^2 U_{ik}
\label{EC}
\ee
with $P=\te{P}{_i^i}$, the cosmological constant $\lambda$, and the additional gravitational potential due to the torsion
\ba
\kappa^2 U_{ik}\!=\!
\te{C}{^j_i_j}\te{C}{^l_k_l}
\!-\! \te{C}{^l_i_j}\te{C}{^j_k_l}
\!-\!{g_{ik}\over 2} 
\!\!\left (\!
\te{C}{^j^m_j}\te{C}{^l_m_l}
\!-\!\te{C}{^m^j_l}\te{C}{^l_j_m}
\!\right ).
\end{align}
Secondly, by variation of (\ref{1}) with respect to the contorsion, the Einstein-Cartan equations connect the torsion tensor to the spin tensor $s$ by
\be
S_{kij}=-{\kappa\over 2} (s_{ijk}-g_{k[j}s_{i]ll})
\label{EC1}
\ee
which provides the relation of the contorsion tensor to the spin itself
\be
C_{kij}={\kappa\over 2} (2 s_{k(ij)}-s_{ijk}+g_{kj}s_{ill}-g_{ij}s_{kll}).
\label{Cs}
\ee
The additional gravitational potential becomes
\ba
U_{ik}=&{1\over 2}\left [
\te{s}{_i^j_j}\te{s}{_k^l_l}
-\te{s}{_i^j_l}\te{s}{_k^l_j}
-\te{s}{_i^j_l}\te{s}{_k_j^l}
+\frac 1 2 \te{s}{^j^l_i}\te{s}{_j_l_k}
+
{g_{ik}\over 2}\left (
\frac 1 2 \te{s}{^j^l^m}\te{s}{_j_l_m}
-\te{s}{^l_j_m}\te{s}{^j^m_l}
+\te{s}{^l_j_l}\te{s}{^j^m_m}
\right )
\right ].
\label{U}
\end{align}

The Einstein tensor as left side of (\ref{EC}) obeys the double contracted Bianchi identity
$\nabla_l^{\{\}}G^{lk}=0$ with respect to Levi-Civita connection which establishes the conservation law
\be
\nabla_l^{\{\}}(T^{lk}+\varepsilon Z^{lk}+\kappa U^{lk})=0
\label{con}
\ee
which has been silently used as consistence check in all forthcoming calculations.

\section{Consistence conditions}
\subsection{Modified Weyssenhoff spin fluid}

The equations are drastically simplified if we work with the Weyssenhoff spin liquid \cite{Ob87} assuming that the spin tensor takes the form $s_{ijk}=s_{ij}u_k$ with the velocity $u$ and the remaining antisymmetric spin tensor being orthogonal $s_{ij} u^j=0$ known as Frenkel condition. Then the additional gravitational potential (\ref{U}) simplifies to
\be
U_{ik}={1\over 2} \left [\te{s}{_i^j}s_{jk}+\sigma^2\left (u_iu_k +{g_{ik}\over 2}\right )
\right ]
\label{U1}
\ee 
with (please note
)
\be
2 \sigma^2=\te{s}{^m_l}\te{s}{_m^l}.
\label{sigma}
\ee
This additional gravitational potential gives a quadratic contribution in $\kappa$ to the Einstein-Cartan equation (\ref{EC}). We do not assume any further averaging and take all linear terms into account.

The contortion tensor (\ref{C}) and torsion tensor become according to (\ref{Cs}) and (\ref{EC1})
\be
C_{kij}&=&{\kappa\over 2}(s_{ki}u_j+s_{kj}u_i+s_{ji} u_k)\nonumber\\
S_{kij}&=&-{\kappa\over 2}s_{ij}u_k
\label{C1}
\ee
and $S_l=\te{S}{^k_i_k}=0$.

\subsection{Condition for the spin tensor}

The additional potential $Z_{ik}$ from the Belinfante-Rosenfeld equation (\ref{BR}) deserves now a closer look. Using (\ref{C1}) we obtain
\be
Z_{ik}={1\over \kappa}\nabla_l\te{C}{_k^l_i}
=
\frac 1 2 \nabla_l (\te{s}{_k^l} u_i+\te{s}{_i^l}u_k+\te{s}{_k_i}u^l).
\label{Z2}
\ee
We see from the Einstein-Cartan equation (\ref{EC}) that the terms should be symmetric in the indices $i,k$. This is visible in (\ref{U1}) but has to be demanded for (\ref{Z2}). This requirement is nothing but the conservation of the spin density (see 2.4.16 of \cite{Pop09}). Applying (\ref{rule}) we translate the covariant derivatives with respect to Levi-Civita connections 
\be
&&Z_{ik}={1\over \kappa}\nabla_l\te{C}{_k^l_i}={1\over \kappa}\nabla_l^{\{\}}\te{C}{_k^l_i}
+{1\over \kappa}\left (
-\te{C}{^m_k_l}\te{C}{_m^l_i}+\te{C}{^l_m_l}\te{C}{_k^m_i}
-\te{C}{^m_i_l}\te{C}{_k^l_m}
\right ).
\label{Z1}
\ee
We need 
\be 
\te{C}{_k^l_i}-\te{C}{_i^l_k}=\kappa s_{ki}u^l
\ee
for the requirement 
\ba
0&=Z_{ik}-Z_{ki}
\nonumber\\
&=\nabla_l^{\{\}}(s_{ki}u^l)
+\te{C}{^m_i_l}s_{mk}u^l
+\te{C}{^m_k_l}s_{im}u^l
+\te{C}{^l_m_l}s_{ki}u^m
\nonumber\\
&=\nabla_l^{\{\}}(s_{ki}u^l)
\nonumber\\
&=\p l (s_{ki}u^l)-\ch{m}{kl}s_{mi}u^l-\ch{m}{il}s_{km}u^l+\ch{l}{ml} s_{ki}u^m
\label{sys}
\end{align}
where we have used (\ref{C1}) and $s_{ij} u^j=0$ as well the antisymmetry of $s$ to see the step from the second to the third line.

We will use a coordinate system where $u^l=(u^0,0,0,0)$. This implies that the antisymmetric spin tensor $s$ has a first zero column and row
\be
s=\begin{pmatrix}
0&0&0&0\cr
0&0&c&-b\cr
0&-c&0&a\cr
0&b&-a&0
\end{pmatrix}
\label{smatrix}
\ee
and (\ref{sys}) translates into 6 equations for $a,b,c$. The first set of linear equations appear for $i=0$  
\be
\ch{m}{00} s_{mk}=0
\ee
which solution shows that we have to have
\be
a=\ch{1}{00} d;\quad b=\ch{2}{00} d;\quad c=\ch{3}{00} d
\label{d}
\ee
with a single unknown function $d$.
The second set of equations appear for $(i=1,k=2), (i=1,k=3), (i=2,k=3)$ and read
\ba
{1\over u^0}\p 0(cu^0)+\ch{3}{01} a+\ch{3}{02} b+\left (\ch{0}{00}+\ch{3}{03}\right ) c=&0
\nonumber\\
{1\over u^0}\p 0(bu^0)+\ch{2}{01} a+\left (\ch{0}{00}+\ch{2}{02}\right )b+\ch{2}{03} c=&0
\nonumber\\
{1\over u^0}\p 0(au^0)+\left (\ch{0}{00}+\ch{1}{01}\right )a+\ch{1}{02} b+\ch{1}{03} c=&0.
\label{d1}
\end{align}
Together with (\ref{d}) these are 3 differential equations for $d$ and the metric tensor $g$. It is interesting to note that the metric is obviously directly dependent on the spin content boiled down to a single function $d$. Later we will consider various solutions of the Einstein-Cartan equations for the metric and will observe the restrictions (\ref{d}) and (\ref{d1}) carefully. Unfortunately, both constraints are mostly not respected with the presented exact solutions in the literature.

\subsection{Einstein-Cartan equations}
Using
\be
\te{C}{_k^l_i}+\te{C}{_i^l_k}=\kappa (\te{s}{_k^l}u_i+\te{s}{_i^l}u_k),
\ee
the contribution of the Belinfante-Rosenfeld equation 
\be
Z_{ik}=\frac 1 2 (Z_{ik}+Z_{ki})=\tilde Z_{ik}+\kappa \tilde U_{ik}
\label{Bel}
\ee
can be seen to split into 
\be
\tilde Z_{ik}=\frac 1 2 \nabla^{\{\}}_l(\te{s}{_k^l}u_i+\te{s}{_i^l}u_k)
\label{Z2a}
\ee  
and a part entering linear in $\kappa$
\be
\tilde U_{ik}=-{1\over 2}(2\sigma^2u_iu_k+\te{s}{_i_l}\te{s}{^l_k}).
\label{tU}
\ee
This splitting corresponds to the two parts in (\ref{Z1}).
Summarizing, the Einstein-Cartan equations (\ref{EC}) take the form
\be
P_{ik}=\left (\lambda\!+\!\frac P 2\right ) g_{ik}\!+\!\kappa ({T}_{ik}\!+\!\varepsilon \tilde Z_{ik})\!+\!\kappa^2 (U_{ik}\!+\!\varepsilon\tilde U_{ik}).
\label{EC1a}
\ee

The canonical energy momentum tensor is assumed to have the form
\be
T_{ik}=(n+p)u_iu_k-pg_{ik}
\ee
with the mass density $n$ and pressure $p$. 

The torsion modifies the Einstein-Cartan equation by an additional energy-momentum tensor part $\tilde Z$ of (\ref{Z2a}) and a potential quadratic in $\kappa$ according to  (\ref{tU}) and (\ref{U1})
\be
U_{ik}\!+\!\varepsilon \tilde U_{ik}\!=\! {\sigma^2\over 2}\left [(1\!-\!2 \varepsilon) u_iu_k\!+\!{g_{ik}\over 2}\right ]\!+\!{1\!-\!\varepsilon\over 2}\te{s}{_i_l}\te{s}{^l_k}.
\ee

\section{Static spherically symmetric solution} 
\subsection{Torsion potentials}
First we investigate the spherically symmetric solution with a diagonal metric
\be
g={\rm diag}\{B(r),-A(r),-R(r),-R(r) \sin^2 \theta\}
\label{gmetric}
\ee
which should include the Schwarzschild solution and generally any static spherically symmetric solutions. We choose freely $u^\mu=(u^0,0,0,0)$ which provides with $u^\mu u_\mu=1$ the form $u^\mu=(\sqrt{B(r)},0,0,0)$ and $u_\mu=(1/\sqrt{B(r)},0,0,0)$.
Calculating the Christoffel symbols, the set of equations (\ref{d}) provides $b=c=0$ and
\be
a={B'(r)\over 2 A(r)} d(r) \sin\theta.
\ee
Since it is static it solves also (\ref{d1}). The $\theta$ dependence remains undetermined and we choose it here such that
\be
\sigma^2(r)={B'(r)^2\over 4 A(r)^2R(r)^2} d(r)^2
\label{sigma2}
 \ee
becomes independent of $\theta$ since this term appears later in the equations and we search for spherical symmetric ones. 
We see that the spin conservation (\ref{sys}) reduces the spin tensor corresponding to the assumed metric.

The contortion tensor can be easily calculated with the nonzero parts
\be
{\kappa a \sqrt{B(r)}\over 2}&=&C_{032}=C_{203}=C_{230}
=
-C_{302}=-C_{023}=-C_{320}
\ee
and 
the linear-$\kappa$ contribution to the Einstein-Cartan equations $\tilde Z$ according to (\ref{Z2a}) via (\ref{Z1}) appears to be zero for this chosen metric. In other words it does not vanish due to averaging but is exactly zero. 

The quadratic-$\kappa$ contribution to the Einstein-Cartan equations (\ref{tU}) reads
\be 
\tilde U_{ik}=-{\sigma^2\over 2}{\rm diag}\{2B(r),0,R(r),R(r) \sin^2{\theta}\}
\ee
and the additional gravitational potential (\ref{U1}) becomes
\be
U_{ik}={\sigma^2\over 4}{\rm diag}\{3B(r),-A(r),R(r),R(r) \sin^2{\theta}\}.
\ee

\subsection{Internal Schwarzschild solution}

Now we solve the Einstein-Cartan equations. 
It turns out that the right-hand side of (\ref{EC1a}) becomes simplified if we use an effective momentum and energy density 
\be
\bar p(r)&=&p(r)+{\lambda\over \kappa}-{e(r)^2\over 4(1-\varepsilon)}
\nonumber\\
\bar n(r)&=&n(r)-{\lambda\over \kappa}+{(3-4 \varepsilon)e(r)^2\over 4 (1-\varepsilon)}.
\label{bnp}
\ee
We will use in the following
\be
e(r)^2=\kappa (1-\varepsilon) \sigma^2(r)
\ee
defining $e(r)$ via (\ref{sigma2}).
If we do not consider $\tilde U_{ik}$ as in most treatments we will have to set $\varepsilon=0$ instead of $\varepsilon=1$.  

The Riemann tensor provides 3 equations since $P_{33}=P_{22}\sin^2\theta$ on both sides. The right hand side of (\ref{EC1a}) together with the conservation law (\ref{con}) becomes then
\be
P_{00}&=&{\kappa\over 2}B(e^2\!+\!\bar n\!+\!3\bar p)
\nonumber\\
P_{11}&=&{\kappa\over 2}A(\bar n\!-\!\bar p\!-\!e^2)
\nonumber\\
P_{22}&=&{\kappa\over 2}R(\bar n\!-\!\bar p)
\nonumber\\
0&=&\nabla_l[T^{lk}+\tilde Z^{lk}+\kappa (U^{lk}+\tilde U^{lk})]
=
(\bar n+\bar p) {B'\over 2 A}+{\bar p'\over V}-{e^2R'\over 2 A R}
\label{cons}
\ee
where we suppress the $r$-dependence of all functions and denote the derivative by $R'$. One can get rid of $B$ combining
\ba
{P_{00}\over 2 B}\!+\!{P_{11}\over 2 A}\!+\!{P_{22}\over R}
=\frac{A' g'}{2 A^2
   R}\!+\!\frac{R'^2\!-\!4 R R''}{4
   A R^2}\!+\!\frac{1}{R}
=\kappa \bar n.
\end{align}
This differential equation is solved as
\be
A(r)={R'(r)^2\over 4\pi R(r)}\left [1-\kappa {m(r)\over 4 \pi \sqrt{R(r)}}\right]^{-1}
\label{A}
\ee
with the total "mass" included
\be
m(r)=2 \pi \int\limits_{r_0}^r d\bar r\sqrt{R(\bar r)} R'(\bar r) \bar n(\bar r)+C_0.
\label{m}
\ee
The integration constant we absorb in $m(r)$ setting $C_0=0$ by assuming a proper $r_0$. 

Next we consider
\ba
{P_{00}\over B}+{P_{11}\over A}={A'R'\over 2A^2R}+{B'R'\over 2 A B R}-{2 R R''-R'^2\over 2 A R^2}=\kappa (\bar n+\bar p).
\label{h1}
\end{align}
Using the conservation (\ref{cons}) we have 
\be
{B'\over B}={e^2R'-2R \bar p'\over R(\bar n+\bar p)}. 
\label{43}
\ee 
With the help of this we eliminate $B$ in (\ref{h1}) and $A$ by (\ref{A}) to obtain the modified Oppenheimer-Volkov equation \cite{BHT16}
\be
\bar p'=-{\kappa R'(\bar n+\bar p)\left (\bar p+{m\over 4 \pi R^{3/2}}\right )\over 4\left (1 -{\kappa \over 4 \pi \sqrt{R}}m\right )}+e^2{R'\over 2 R}.
\label{OV}
\ee
Using this in (\ref{43}) we obtain another form for $B'/B$ 
\be
{B'\over B}={\kappa R'\over 8\pi R^{3/2}}\,{m+4 \pi R^{3/2} \bar p\over 1-{\kappa\over 4 \pi \sqrt{R}} m}.
\label{Bs}
\ee

With these solutions at hand we check that the last equation
\be
{P_{22}\over R}={1\over R}+{A'R'\over 4 A^2R}-{B'R'\over 4 A B R}-{R''\over 2 A R}
={\kappa\over 2}(\bar n-\bar p)
\ee
of the Einstein-Cartan equations (\ref{cons}) is completed identically.
The equations (\ref{Bs}), (\ref{OV}) and (\ref{A}) solve therefore the Einstein-Cartan equations together with the conservation law (\ref{cons}) and are the general solutions of the assumed spherical-symmetric and static metric (\ref{gmetric}) within the Weyssenhoff fluid. The known internal Schwarzschild solution is visible for $e\to 0$ and $\lambda \to 0$. Of course, the function $R(r)$ as prefactor of the angular parts in the metric remains undetermined dependent on the used coordinate system. A simple variable transformation $R(r)={\bar r}^2$ would fix it in the standard form. Further treatments can be performed numerically dependent on the given momentum and density profile via the equation of state. Exact solutions for the inner region of compact objects are discussed in \cite{CZJ18,LS19,DN19}, for rotating stars in \cite{Gal09} and possible upper limits of masses of stars in \cite{BHT16}. 

An important remark we have to make. The so far undetermined torsion $e(r)$ cannot be chosen arbitrarily since (\ref{d1}) and the Oppenheimer-Volkov equation (\ref{OV}) provides with (\ref{bnp}) an internal consistence even out of matter as we will see now.

\subsection{External Schwarzschild solution \label{SCHW}}
 In the exterior of stars we can set $p=0$ and $n=0$ and should obtain the external Schwarzschild solution. The effective mass (\ref{m}) and the Oppenheimer-Volkov equation (\ref{OV}) read then
\ba
m'&=\pi \sqrt{R}R' \left (-{2 \lambda \over \kappa}+{(3-4 \varepsilon) e^2\over 2 (1-\varepsilon)}\right )
\nonumber\\
{(\e^2)'\over e^2}&=R' {{(5-6 \varepsilon)\kappa m\over 8\pi R^{3/3}}- {2 (1-\varepsilon)\over R(r)}-\frac {1-2 \varepsilon}{2} \lambda-{(1-2 \varepsilon)\over 8(1-\varepsilon)}\kappa e^2(r) \over 1-{\kappa m(r)\over 4 \pi \sqrt{R(r)}}} .
\label{47}
\end{align}
Please note that even in the matter-free space $n=0, p=0$ we do not have a constant $R(r)$ and therefore $m(r)$. Due to the allowed torsion $e(r)$ the function $m(r)$ is loosing its meaning as mass parameter outside the stars and is an auxiliary function of $r$. Not setting it constant as in the normal exterior Schwarzschild solution is only justified if it solves the Einstein-Cartan equation what we will show now.

The first equation of (\ref{47}) provides $e=e(m')$ and inserted into the second leads to a second-order nonlinear differential equation for $m(r)$. Somewhat simplified one obtains with the transformation $R(r)=r^2$ and
\ba
e^2(r)&={4(1-\varepsilon)\over \kappa (3-4 \varepsilon)}\left ({1\over r^2}+\lambda +{\bar m'(r)\over r^2}\right )\nonumber\\
m(r)&={4 \pi \over \kappa} [r+\bar m(r)],
\end{align}
the equation for $\bar m(r)$
\ba
&r (1-2 \varepsilon) \left (\bar{m}'+\lambda r^2+1\right) \left [\bar{m}'-(2-4 \varepsilon)
   (\lambda r^2+1)\right]
\nonumber\\
&
=(3-4 \varepsilon) \bar{m}
   \left[r\, \bar{m}''+(3-6 \varepsilon) (\bar{m}'+1)+\lambda r^2 (5-6 \varepsilon)\right].
\label{dgl}
\end{align}
A series ansatz shows that this equation has two solutions
\ba
\bar m(r)=m_0{\kappa\over 4 \pi}-r-{\lambda\over 3} r^3,\quad \bar m(r)=-r  \left ({\lambda\over 3} r^2+{1-2 \varepsilon\over 4-6\varepsilon}\right ) 
\end{align}
where the constant $m_0$ in the first solution will turn out to be the mass of a star. Translated back we obtain with (\ref{A}) and (\ref{Bs}) two solutions. The first one
\be
A(r)&=&{{R'(r)^2\over 4R(r)}\over 1-{\kappa m_0\over 4\pi \sqrt{R(r)}}+{\lambda \over 3} R(r)}
\nonumber\\
B(r)&=&C_1\left (1-{\kappa m_0\over 4\pi \sqrt{R(r)}}+{\lambda \over 3}R(r)\right )
\nonumber\\
e(r)&=&0
\nonumber\\
 m(r)&=&m_0+{4\pi \lambda\over 3 \kappa} R(r)^{3/2}
\ee
is the standard Schwarzschild solution with zero torsion and the extension to include the cosmological constant known as Kottler solution \cite{CXZ87}. Actually, for large distances $R(r)\to \infty$ the factor $B\to 1+2\phi(r)$ should approach the gravitational potential $\phi(r)=-{\kappa m_0/ 8 \pi\sqrt{R(r)}}$ and consequently $C_1=1$. The standard coordinate system appears of course for $R(r)=r^2$.

Abbreviating
\be
C={3 (1-2 \varepsilon)\over 2(2-3\varepsilon)}
\label{Con}
\ee
the second solution with finite torsion
\be
A(r)&=&{3 R'(r)^2\over 4 R(r)[C +{\lambda }R(r)]}
\nonumber\\
B(r)&=&C_1 R(r)
\nonumber\\
\sigma^2(r)&=&{e(r)^2\over \kappa(1-\varepsilon)}={2\over (2-3\varepsilon)\kappa^2 R(r)}
\nonumber\\
 m(r)&=&{4\pi\over 3 \kappa} \sqrt{R(r)}\, \left [3-C-\lambda R(r)\right ]
\label{sol2}
\ee
has unusual properties which will be explored in the following chapters. 
At a first sight  
in order to show the gravitational potential in $B(r)$ it seems to require $R(r)= [1-2\phi(r)]/C_1$ for large $r$ and it would mean
\be
B(r)&=&1-{\kappa m_0\over 4 \pi r}=C_1 R(r)
\nonumber\\
A(r)&=&{{3 \kappa^2 m_0^2\over 64 \pi^2 r^4}\over \left (1-{\kappa m_0\over 4\pi r}\right )\left [\lambda (1-{\kappa m_0\over 4\pi r})+ C_1 C\right ]
} 
=
{3 \kappa^2 m_0^2\over 64 \pi^2 r^4}{1 \over \lambda + C_1 C
}+o(r^{-5})
\nonumber\\
\sigma^2(r)&=&{e(r)^2\over \kappa(1-\varepsilon)}={2 C_1\over \kappa^2(2-3 \varepsilon)} {1\over 1-{\kappa m_0\over 4 \pi r}}
= {2C_1\over \kappa^2(2-3 \varepsilon)}+o(r^{-1}).
\label{53}
\ee
We see that this solution seemingly creates a conflict between the demand of gravitational potential $B$ at large distances and the asymptotic flatness. We will discuss below in chapter~\ref{discuss} an appropriate coordinate system to discuss this question more deeply. 

In the standard coordinate system
$R(r)=r^2$ we have for the metric (\ref{sol2})
\be
A(r)&=&{3\over \lambda r^2\!+\!C}={1\over 1\!-\!{b(r)\over r}};\quad b(r)=r\!-\!\frac 2 3 r^2(C\!+\!\lambda r^2)
\nonumber\\
B(r)&=&C_1 r^2
\nonumber\\
\sigma^2(r)&=&{e(r)^2\over \kappa(1-\varepsilon)}={2\over (2-3\varepsilon) \kappa^2 r^2}.
\label{stand}
\ee

In the next chapter we will see what modifications can appear allowing a non-diagonal metric. Especially we will prove that our series ansatz solution is indeed the only one besides the torsion-free Schwarzschild solution and will find appropriate coordinate systems where the physics becomes transparent. 

This second solution {\ref{stand}) or (\ref{sol2}) does not provide a parametrically change of the Schwarzschild solution with respect to torsion. The only remaining torsion parameter $\sigma$ itself becomes fixed by the Einstein-Cartan equations itself and is proportional to the squared inverse coupling constant similar to bimetric gravity in first order \cite{DN19}. This is here fundamentally different from solving the equations in matter where the torsion parameter are related to spin. In free space we see that a second solution appears with completely determined metric and torsion just by allowing the fact that torsion is present in space-time. We will discuss the implication of this solution below in chapter~\ref{discuss}. 

\subsection{Comparison with known solutions}
In order to compare with the solution in the literature let us specify some limits of  (\ref{sol2}). It is sufficient to compare specific angular trajectories in order to demonstrate the difference to known metrics. We will use spherical symmetric coordinates and choose specifically $d\Omega^2=0$ to compare the parts in time and radial coordinates. 

In \cite{Ka10} a solution with $B(r)=r^4$ has been found which would mean $R(r)=r^4/C_1$ according to (\ref{sol2}) and our metric (\ref{gmetric}) would become
\be
A(r)&=&{12r^2\over \lambda r^4+C_1 C}
\nonumber\\
B(r)&=&r^4
\nonumber\\
R(r)&=&\frac{r^4}{C_1}.
\nonumber\\
\sigma^2(r)&=&{e(r)^2\over \kappa(1-\varepsilon)}={2 C_1\over \kappa^2(2-3\varepsilon) r^4}.
\ee
This does not reproduce eq 6.7. of \cite{Ka10}. It is noted that this solution \cite{Ka10} itself does not complete the Einstein equation seen by direct insertion. The more general solution in \cite{Ka19} with $B(r)=r^{2n}$ would mean here
\be
A(r)={3 n^2\over r^2\left (\lambda +{c\over r^{2n}}\right )}
\ee
again different from the one in Eq 5.9 of \cite{Ka19}.

With Eq. 159 in \cite{LS19}, a solution was presented for inner stars with junction to the outside leading to $B(r)=(a+b r^2)^2$ which would mean with our solution 
\be
A(r)={12 b^2 r^2\over C+\lambda (a+b r^2)^2}
\ee
which is different from the one published there $A(r)=1/(1+c r^2/(a+3b^2)^{2/3})$.

Our solution here is a class-A type of \cite{SZH18} though by itself different as one sees in Eq. 41 where the authors give $B(r)=(c_2+c_1/\sqrt{r})^2$. This would lead in our solution to
\be
A(r)=F^{-1}={{3\over r^2}\over 4\lambda(1+{c_2\over c_1\sqrt{r}})+{C+c_2^2\lambda\over c_1^2}r}
\ee
distinct from $F$ presented there. Similarly in \cite{CZJ18} it was given
$B(r)=1-2m/r$ which would translate in our solution to
\be
A(r)={3 m^2\over r^2(r-2m)[C r+\lambda(r-2m)]}
\ee
which is distinct from Eq. 16 of \cite{CZJ18} which reads
\be
A(r)=F^{-1}={(r-2 m)(2 r+a)\over r(2 r-3m)}
\ee
even without cosmological constant.

In \cite{ZMRS14} the authors considers equations of motions resulting from quartic terms of contortion tensor in the Lagrangian which leads to quartic terms of coupling constant in the Einstein-Cartan equations. Consequently the exact solutions outside is different from the one presented here which is quadratic in the coupling constant.

A time-dependent spherically symmetric solution has been obtained in \cite{FGLMY12} with its static configuration (Eq. 79 there) by $A(r)=1/(1+c_1/r^2)$ which would mean for the solution here
\be
B(r)={C\over \lambda}\!\sinh^2\!\sqrt{{\lambda\over 3}(c_1\!+\!r^2)}\to {C\over 3}(c_1\!+\!r^2)\,{\rm for}\, \lambda=0
\ee
being clearly different from the form $B(r)=1$ in \cite{FGLMY12}. The starting point there was also slightly different as being the Einstein-Cartan-Kalb-Raimond coupling. 

In \cite{MZH17,MZH19} the special wormhole form of Morris and Thorne \cite{MT88,MTY88}
\be
ds^2={\rm e}^{2\Phi(r)}dt^2-a(t)\left ( {dr^2\over 1+{b(r)\over r}}+r^2d\Omega^2\right )
\ee
has been employed with $\Phi(r)=0$. From our standard form (\ref{stand}) we see $\Phi(r)=\ln r$ and therefore a different form.

If we compare with the static solution $A(r)=a/(1-b_0/r)$ we would obtain for our solution
\be
B(r)=-{C\over \lambda} \sin^2\!\left [ \!\sqrt{\lambda \over 3} \left (\!\sqrt{r(b_0\!-\!r)}\!-\!b_0\,{\rm arcsin}\sqrt{r\over b_0}\right )\right ]
\ee
clearly different from $B(r)=1$ reported there. 

An exact solution of axially symmetric solution was presented in \cite{MSS16} 
with $B(t)=\sqrt{c_2/(t+c)}$ and 
\be
A(t,r)=\sqrt{t+c\over c_1}{\rm e}^{2\mu r^2\sin^2\theta}
\ee
which clearly differs from our $B-A$ relation even when rewritten in spherical coordinates.

Therefore we consider (\ref{sol2}) as a not yet reported solution. Since it was obtained by a series ansatz, we have still to prove the uniqueness of this solution which we will see now. Let us remark here that most presented exact solutions do not respect the spin-consistence relation (\ref{d}) and (\ref{d1}).

\section{Static non-spherical solution}

First we test a non-spherical metric in the variables $(t,r,\theta,\phi)$
where we assume additionally to the diagonal elements (\ref{gmetric}) the term
\be
g=\begin{pmatrix}
B(r)&0&g_{02}(r)&0\cr
0&-A(r)& 0&0\cr
g_{02}(r)&0&-R(r)&0\cr
0&0&0&-R(r) \sin^2{\theta}
\end{pmatrix}.
\label{m1}
\ee
Then the spin-conservation (\ref{d1}) yields the condition $b=c=0$ and
\be
a(t,r,\theta) B'(r) [B(r) g'_{02}(r)-g_{02}(r) B'(r)]=0.
\ee
For nonzero torsion this demands 
\be
g_{02}(r)=g_{02} B(r). 
\ee
The case of constant $B(r)$ will be seen to be included in the solution below.

Now lets solve the Einstein-Cartan equations in the coordinate system $(t,r,\theta,\phi)$. The linear-$\kappa$ contribution to the Einstein-Cartan equations $\tilde Z$ according to (\ref{Z2a}) via (\ref{Z1}) are still exactly zero. It becomes nonzero for time-dependent metrics. The nonzero components of the potential (\ref{tU}) read
\ba 
&\tilde U_{00}=-\sigma^2 B(r),\,\tilde U_{02}=\tilde U_{20}=-\sigma^2g_{02} B(r),
\nonumber\\
& \tilde U_{22}=-{\sigma^2\over 2}[3 g_{02}^2 B(r)+R(r)],\, \tilde U_{33}=-{\sigma^2\over 2}R(r)\sin^2\theta
\end{align}
and the additional gravitational potential (\ref{U1}) becomes
\ba
&U_{00}=\frac 3 4 \sigma^2 B(r),\,U_{02}=-\tilde U_{20}=\frac 3 4 \sigma^2g_{02} B(r),
\nonumber\\
& U_{11}=-{\sigma^2\over 4}A(r),\, U_{22}={\sigma^2\over 4}[4 g_{02}^2 B(r)+R(r)],
\nonumber\\
& U_{33}={\sigma^2\over 4}R(r)\sin^2\theta.
\end{align}

We search now immediately for the free- space solution without matter $n=0, p=0$. Then it is revealing to consider first the $(23)$ element $G_{23}$ of the Einstein-Cartan equations (\ref{EC1a}) which provides
\be
P_{23}=\frac{g_{02}^2 \cot\theta \left[R(r) B'(r)-B(r) R'(r)\right]}{2 R(r)
   \left[g02^2 B(r)+R(r)\right]}=0
\ee
demanding 
\be
B(r)=C_1 R(r).
\label{sB}
\ee 
Next we consider the conservation (\ref{con}) which becomes
\be
\left \{ 0,-{R(r) e^2(r)'+e^2(r) R'(r)\over4(1-\varepsilon)A(r)(1+g_{02}^2C_1) R(r)},0,0\right \}=0
\ee
and which is fulfilled for 
\be
e^2(r)={C_2\over R(r)}.
\label{se}
\ee
With these two results for $B(r)$ and $e(r)$ we obtain from
\be
P_{11}={3 A' R'\over 4 A R}-{6 R R''-3(R')^2\over 4 R^2}=-\lambda A(r)
\ee
the solution
\be
A(r)={3 (R'(r))^2\over 3 C_3 R(r)+4 \lambda R(r)^2}.
\label{sA}
\ee 
With the help of (\ref{sB}), (\ref{se}), (\ref{sA}) we get
\be
P_{02}&=&{g_{02}\over 2} C_1(C_3+2 \lambda R(r))
=
{g_{02}\over 2} C_1\left [
{\kappa (1-2 \varepsilon) C_2\over (1-\varepsilon)(1+g_{02}^2 C_1}+2\lambda R(r)
\right ]
\ee
which provides the constant
\be
C_2={(1-\varepsilon)(1+g_{02}^2 C_1) C_3\over \kappa (1-2 \varepsilon)}.
\label{sC2}
\ee
The possibility $C_1=0$ we exclude since this would mean $B(r)=0$ according to (\ref{sB}).
Finally we consider with (\ref{sB}), (\ref{se}), (\ref{sA}), (\ref{sC2})
\be
P_{22}&=&1-\frac 1 2 C_3-\lambda R(r)
=
C_3{1-\varepsilon+(2-3 \varepsilon) C[1]g_{02}^2\over 2 (1-2 \varepsilon)}-\lambda R(r)
\ee
which provides the constant
\be
C_3={2(1-2 \varepsilon)\over (2-3 \varepsilon) (1+g_{02}^2 C_1)}.
\label{sC3}
\ee
The solutions (\ref{sB}), (\ref{se}), (\ref{sA}), (\ref{sC2}) and (\ref{sC3}) fulfill the Einstein-Cartan equations and the conservation laws.
Collecting together we have the metric (\ref{m1}) with 
\be
A(r)&=&{3 R'(r)^2\over 4 R(r)[{C\over  1+g_{02}^2 C_1}+\lambda R(r)]}
\nonumber\\
B(r)&=&C_1 R(r)
\nonumber\\
g_{02}(r)&=&g_{02} C_1 R(r)
\nonumber\\
\sigma^2(r)&=&{e(r)^2\over \kappa (1-\varepsilon)(1+g_{02}^2 C_1)}
=
{2\over \kappa^2 (2-3 \varepsilon) (1+g_{02}^2 C_1) R(r)}
\nonumber\\
a(r,\theta)&=&\sqrt{2 R(r)\over 2-3 \varepsilon}{\sin \theta\over \kappa}.
\label{sol3}
\ee
In the limit of spherical symmetry, $g_{02}\to 0$ we recover exactly the solution (\ref{sol2}). Due to the unique way of deduction we have proven in this way that this is the only solution of the complicated differential equation (\ref{dgl}) besides the torsion-free Schwarzschild solution. Still the concern remains that there is no smooth transition from vanishing torsion $\sigma^2\to 0$ towards the Schwarzschild solution.

\subsection{Uniqueness of non-spherical second solution}
We will now test successively further off-diagonals in the metric to show that the metric (\ref{m1}) with (\ref{sol3}) is indeed a unique second solution besides the Schwarzschild one.

First we consider the metric
\be
g=\begin{pmatrix}
B(r)&0&g_{02}(r)&0\cr
0&-A(r)& g_{12}(r)&0\cr
g_{02}(r)&g_{12}(r)&-R(r)&0\cr
0&0&0&-R(r) \sin^2{\theta}
\end{pmatrix}.
\label{m2}
\ee
Then the spin-conservation (\ref{d1}) yields the condition $c=0$ and
$g_{02}(r)=C_1 B(r)$ as in the foregoing chapter. Working it through further one considers the $(0,2)$ equation of the Einstein-Cartan equation which provides
\ba
{g_{02} g_{12}(r) B'(r)\over 2 A(r)[G_{02}^2 B(r)\!+\!R(r)]\!-\!2 g_{12}^2(r)}\cot\theta+...=0
\end{align}
as only angular dependence. This means either $g_{12}(r)=0$ and the solution as before or alternatively $g_{02}=0$. This last option is excluded since the component $(1,2)$ yields
\ba
&{g_{12}^2(r)R'(r)+g_{02}^2 A(r)(R(r) B'(r)-B(r) R'(r))\over 2 R(r) [ A(r) (g_{02}^2 B(r)+R(r))-g_{12}^2(r)]}\cot\theta+...
=0
\end{align}
which would lead also to $g_{12}(r)=0$ if $g_{02}=0$. We conclude that the Einstein-Cartan equations require for the metric (\ref{m2}) that the latter condition holds and we are back to (\ref{m1}).

Next, we test the metric
\be
g=\begin{pmatrix}
B(r)&0&g_{02}(r)&0\cr
0&-A(r)& 0&g_{13}(r)\sin\theta\cr
g_{02}(r)&0&-R(r)&0\cr
0&g_{13}(r)\sin\theta&0&-R(r) \sin^2{\theta}
\end{pmatrix}
\label{m3}
\ee
where the $\sin \theta$ is chosen such that the determinant and spin components are separable in angular variables.
Then the spin-conservation (\ref{d1}) yields the condition $b=0$ and
$g_{02}(r)=C_1 B(r)$. Considering then the $(2,2)$ component of the Einstein-Cartan equation shows
\be
{g_{13}(r)^2\over 2(g_{13}^2(r)-A(r) R(r))}{1\over \sin^2\theta}+...=0
\ee
demanding $g_{13}=0$ and again the former solution remains.

Finally we test the metric
\be
g=\begin{pmatrix}
B(r)&0&g_{02}(r)&0\cr
0&-A(r)& g_{12}(r)&g_{13}(r)\sin\theta\cr
g_{02}(r)&g_{12}(r)&-R(r)&0\cr
0&g_{13}(r)\sin\theta&0&-R(r) \sin^2{\theta}
\end{pmatrix}
\label{m4}
\ee
Then the spin-conservation (\ref{d1}) yields the condition $b=0$ and
$g_{02}(r)=C_1 B(r)$. Considering then the $(3,3)$ component of the Einstein-Cartan equation shows
\ba
&{A(r) g_{13}^2(r)\over 2 [A(r) R(r)\!-\! g_{13}(r)^2] [g_{02}^2 B(r)\!+\!R(r)]\!-\!2 R(r) g_{12}^2(r)}
{\cot^2\theta}+...\cot\theta+...=0
\end{align}
which yields $g_{13}=0$. Then we consider the $(0,2)$ component
\be
{g_{02} g_{12}(r) B'(r)\over 2 A(r) (g_{02}^2B(r)+R(r))-2 g_{12}(r)^2}\cot\theta+...=0
\ee
which demands either $g_{12}=0$ and we have the foregoing case or $B(r)=const$. The latter case leads for the $(0,2)$ component to
\be
- B g_{02} \lambda=0
\ee
which means $g_{02}=0$. But then we get for $(1,2)$ component
\be
{g_{12}^2(r) R'(r)\over 2 A(r) R^2(r)-2 R(r) g_{12}^2(r)}=0
\ee
which demands $g_{12}(r)=0$ and again the former solution remains.

The general metric consists of 6 functions to be determined by the Einstein-Cartan equations since 4 conditions can be fixed by coordinate transformations. 
Since the metric (\ref{m4}) consists of 6 unknown functions we consider them as the general solution. As discussed the Einstein-Cartan equations boil all them back to the solution (\ref{m1}) with (\ref{sol3}) which we consider therefore as the unique second solution besides the Schwarzschild one outside matter as discussed in chapter~\ref{SCHW} above. This statement relates to the first-order action (\ref{1}) in the Einstein-Hilbert scheme and not to any other forms. 

\section{Discussion of the second solution\label{discuss}}

Besides the torsion-free Schwarzschild solution in free space we found a unique second torsion solution (\ref{m1}) with (\ref{sol3}) and $g_{02}(r)=g_{02}B(r)=C_1 g_{02} R(r)$. Without loss of generality we can use the standard $R(r)=r^2$ in the following. The metric reads then
\be
ds^2&=&C_1 r^2 d t^2-{3\over {C\over 1+C_1 g_{02}^2}+\lambda r^2} d r^2
-r^2(d\theta^2+2 C_1 g_{02} d\theta d t+\sin^2\theta d\phi^2).
\label{met}
\ee
The off-diagonal term $\sim g_{02}$ creates a rotation in the azimuthal angle with the angular velocity $\Omega^2=1/(1+1/C_1 g_{02}^2)$ as one can see using the transformation
\be
\bar t&=&\sqrt{C_1 (1+C_1 g_{02}^2)} t
\nonumber\\
\bar \theta&=&\theta+\Omega \bar t
\ee
which leads to
\ba
ds^2=r^2 d\bar t^2\!-\!{3 d r^2\over {C\over 1\!+\!C_1g_{02}^2}\!+\!\lambda r^2} 
\!-\!r^2[d\theta^2\!+\!\sin^2{(\bar \theta-\Omega \bar t)} d \phi^2].
\end{align}
This was expected since the off-diagonal term at the place $(0,2)$ in the metric tensor has survived as only possible solution and it creates exactly time-angle correlations. 

Another coordinate transformation in time and angles
\be
\hat t&=&\sqrt{C_1\over 1+C_1 g_{02}^2}(t -g_{02} \, \theta)
\nonumber\\
\hat r &=& r \sqrt{1+C_1 g_{02}^2}
\nonumber\\
\hat \phi&=&{\phi\over \sqrt{1+C_1 g_{02}^2}}
\ee
brings the metric (\ref{met}) into the diagonal form
\ba
ds^2&=
\hat r^2 d\hat t^2\!-\!{3 \over C\!+\! \lambda \hat r^2} d \hat r^2
\!-\!\hat r^2(d\theta^2\!+\!\sin^2\theta d\hat \phi^2).
\label{metricd}
\end{align}
We see that the off-diagonal element $g_{02}$ can be transformed away.
Further we will discuss this form (\ref{metricd}).

\subsection{Wormhole} 

For stationary $dt=0$ and equatorial slice $\theta=\pi/2$, the metric (\ref{metricd}) becomes
\be
ds^2=-{3 \over C + \lambda \hat r^2} d\hat r^2-\hat r^2 d\hat \phi^2.
\label{red}
\ee
Please note that the variables can be scaled here by $\hat r\to c \hat r$ and $\hat \phi\to \hat \phi/c$ to result into $C\to c^2 C$. This shows that $C$ appears as an arbitrary constant instead of the originally introduced (\ref{Con}). This is important since the further discussion is dependent on $C$ which is therefore a free parameter irrespective of the Belinfante-Rosenfeld equation which we have tracked by $\varepsilon$ and (\ref{Con}).

We can embed (\ref{red}) in flat 3D cylindrical coordinates \cite{MT88,MTY88,Mu04}
\be
ds^2=-dz^2-d\hat r^2-\hat r^2d\phi^2
\ee
which provides
\be
z(r)&=&\pm\int d \hat r \sqrt{3-C-\lambda \hat r^2\over C+\lambda \hat r^2}.
\ee
We see that for $\lambda<0$ we have a lower and an upper limit for $\hat r$ in order to render $z(\hat r)$ real,
\be
{C-3\over |\lambda|}<\hat r^2<{C\over |\lambda|}.
\label{rC}
\ee
The upper limits gives the finite value of the total universe and 
the throat of the wormhole since both solution touch at this point. Indeed, integrating one gets
\be
z(\hat r)=
 \sqrt{C-3\over \lambda}\, {\rm E}\left ({\rm arcsin}\sqrt{-\lambda \hat r^2\over C}|{C\over C-3}\right )-z_0
\label{zvr}
\ee
with the elliptic E-function ${\rm E}(n|m)=\int_0^nd t\sqrt{1-m\sin^2t} $ and $z_0$ equals the lower possible value of $r$ according to (\ref{rC}) which is zero for $C<3$ .  The results are plotted in figure~\ref{worm} and shows that we have a closed universe for $\lambda<0$ with the upper radius (\ref{rC}). The lower radius appears only for $C>3$ at which the derivative $z'(r)$ diverges and which corresponds to the throat of the wormhole \cite{MT88}. Therefore we see that in the case $C>3$ we do have a wormhole for negative cosmological constant. 

For positive cosmological constant, $\lambda>0$ we obtain (\ref{zvr}) with also an upper boundary for the radius, $\hat r^2<(3-C)/\lambda$, but no lower boundary radius since the formal $-C/\lambda$ one is smaller zero. One sees that no wormhole is possible since the two branches are touching at zero. This universe is possible only for $C<3$ since otherwise the upper boundary would be smaller zero. 

For vanishing cosmological constant there is no upper limit of the radius but $C<3$ 
and the figure is a cone with $z(r)=\pm \sqrt{3-C\over C} r$.

\begin{figure}
\centerline{\includegraphics[width=14cm]{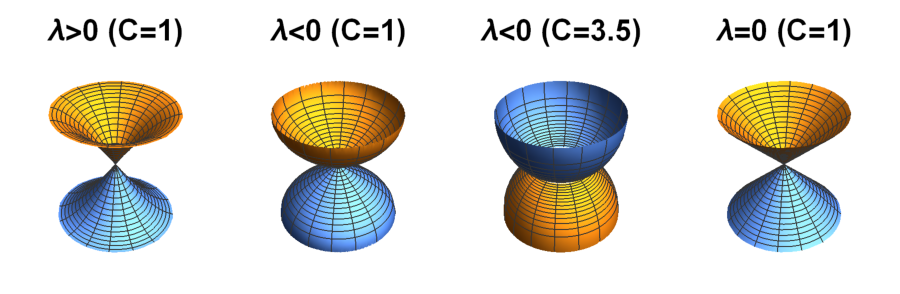}}
\caption{\label{worm} The 3D embedding diagram of the solution with torsion for four different constants $C$ and different cosmological constants. In case of $\lambda<0$ and $C>3$ one gets a finite throat.}  
\end{figure}

\subsection{Accelerations and effects of curvature}

Now we discuss the metric of the second solution (\ref{m1}) with (\ref{sol3}) with respect to possible curvature effects.
The radial acceleration of a particle at rest in this coordinate system is given by the affine geodesics
\be
{d^2 x^1\over d\tau^2}=-\te{\Gamma}{_0_0^1}(u^0)^2=-\frac 1 2 {B'\over A B}={2(C+\lambda R)\over 3 R'}
\ee
and we see that for negative $\lambda$ there is a maximal $R_m=r_m^2=-C/\lambda$ where the acceleration vanishes which corresponds to the upper limit as we have seen in the wormhole discussion in the last chapter. This appears also if we consider light moving radially, $ds^2=d\Omega^2=0$, leading to
\be
\left ({d r\over d t}\right )^2={B\over A}={4 C_1 R^2\over 3 R'^2}(C+\lambda R) .
\ee
In the case $\lambda<0$ there is no light permitted for $R=r^2>R_m=-C/\lambda$ and the universe is closed in this case.

The effect of curvature in the three cases for $\lambda$ is best seen calculating the distance $D$ from a given point in the three-dimensional subspace and the corresponding sphere spanned by all points with this same distance. 

The distance according to geodesics in the three-dimensional subspace is
\ba
D&=\int\limits_0^r d r \sqrt{-g_{11}}=\sqrt{3} 
\left \{
\begin{array}{ll} 
{1\over \sqrt{|\lambda|}} {\rm arcsin}\sqrt{|\lambda| R\over C} &\lambda <0
\cr
{1\over \sqrt{|\lambda|}} {\rm arcsinh}\sqrt{|\lambda| R\over C} &\lambda >0
\cr
\sqrt{R\over C}&\lambda=0
 \end{array}
\right .
\nonumber\\
&
\approx
\sqrt{ 3\over C} \left [\sqrt {R}\pm {|\lambda|\over 6 C}R^{3/2}+o\left ({|\lambda R|\over C}\right)^{5/2}  \right ].
\label{dist}
\end{align}

The area spanned by all points with the same distance becomes
\ba
A&=\int\limits_0^\pi d\theta \int\limits_0^{2\pi} d \phi\sqrt{g_{22}g_{33}}=4\pi R
\nonumber\\
{A\over4\pi D^2}&= 
\!\left \{\!
\begin{array}{ll} 
{{|\lambda| R} \over 3{\rm arcsin}^2\sqrt{|\lambda| R\over C}} &\lambda <0
\cr
{{|\lambda| R}\over 3 {\rm arcsinh}^2\sqrt{|\lambda| R\over c}} &\lambda >0
\cr
 {C\over 3}&\lambda=0
 \end{array}
\!
\right \}\approx {C\over 3} \!+\!o\left (\!{|\lambda| R\over C}\right ).
\end{align}
We see that even at small distances the permanent curvature of $C/3$
is present to deviate from $4\pi$. 
The plots are given in figure~\ref{ad}.  

\begin{figure}
\centerline{
\includegraphics[width=8cm]{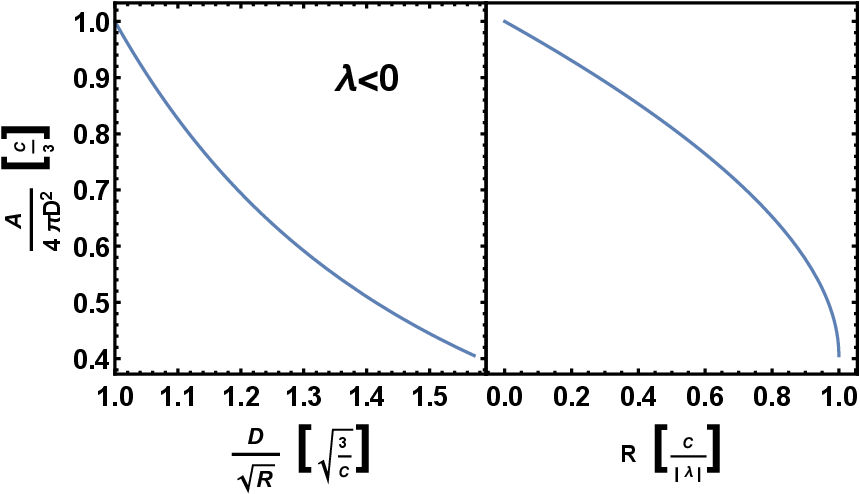}
\includegraphics[width=8cm]{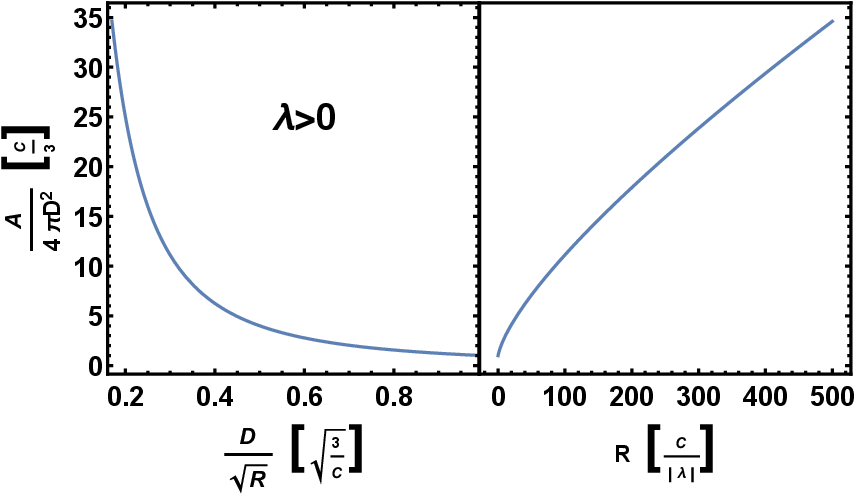}}
\caption{\label{ad} The area spanned in embedded three-dimensional space vs the distance $D$ of (\ref{dist}) (left) and vs coordinate $R$ (right) for $\lambda<0$ 
and $\lambda>0$ 
.}  
\end{figure}

The analogous calculation for the space content reads
\ba
V(r)&=\int\limits_0^r d r\int\limits_0^\pi d\theta \int\limits_0^{2\pi} d \phi\sqrt{-g_{11}g_{22}g_{33}}
\nonumber\\
&={4 \pi\over 3}\sqrt{3\over C} 
\!\left \{\!
\begin{array}{ll} 
{C\over 2 |\lambda|}
\left [{\rm arcsin} \sqrt{|\lambda| R\over C}-\sqrt{R(1-{|\lambda| R\over C})}\right ] &\lambda <0
\cr
&\cr
{C\over 2 |\lambda|}
\left [\sqrt{R(1+{|\lambda| R\over C})}-{\rm arcsinh} \sqrt{|\lambda| R\over C}\right ] &\lambda >0
\cr
&
\cr
 R^{3/2}&\lambda=0
 \end{array}
\!
\right \}
\end{align}
and we see that in the case $\lambda<0$ the space is finite due to the upper limit $R_m=-C/\lambda$ with
\be
V\left (\sqrt{R_m}\right )=\pi^2{\sqrt{3} C\over |\lambda|^{3/2}}.
\ee
For small distances we obtain
\be
{V\over {4\pi \over 3}D^3}\approx {C\over 3} \pm {|\lambda| R\over 15 C}\!+\!o\left (\!{|\lambda| R\over C}\right )^2.
\ee
which again provides the information that there is a permanent curvature even at small distances of $C/3$.

\subsection{Other transformations}

\subsubsection{FLRW forms}

Since we have the freedom to use any coordinate transformation, we might search now for a coordinate system which reveals the most similarity with known metrics. Let us go back to the metric (\ref{metricd}). First we consider the case $\lambda=0$. Using
\be
y=\sqrt{3\over C}\cosh \sqrt{C\over 3}t,\quad 
y=\sqrt{3\over C}\sinh \sqrt{C\over 3}r
\ee
we get
\be
ds^2=dx^2-\left [dy^2+{C\over 3}(y^2-x^2)d\Omega^2\right ]
\ee
which shows a time (x) and space (y) dependent scale. The asymptotic behaviour is seen by the coordinate transformation $\tilde t=\hat r \hat t$ and $r^{\sqrt{C/3}}=\hat r$ which brings the metric (\ref{metricd}) into the form
\be
ds^2&=&d\tilde t^2-{2\over r} \sqrt{C\over 3} d\tilde t dr
-r^{2(\sqrt{C\over 3}-1)}\left [
\left (1+{C \tilde t^2\over 3 r^{2\sqrt{C\over 3}}}\right ) dr^2+r^2d\Omega^2\right ]
\nonumber\\
&\approx&
 d\tilde t^2-r^{2(\sqrt{C\over 3}-1)}\left (
dr^2+r^2d\Omega^2+o({1\over r^2})\right )
\ee
where the last line is for large $r$. We see that the metric becomes asymptotically flat if $C=3$. Otherwise we obtain an asymptotically increasing or decreasing scale factor of the three-dimensional volume.

Next we consider the case $\lambda\gtrless 0$ and scale (\ref{metricd}) by  $\hat r=\sqrt{C/|\lambda|} \bar r$ and $\hat t=\sqrt{3/C} \bar t$ to obtain
\ba
ds^2&=
{3\over |\lambda|}\left [ \bar r^2 d \bar t^2\!-\!{1\over 1\!\pm\! \bar r^2} d \bar r^2
\!-\!{C\over 3}\bar r^2d\Omega^2 \right ]
\label{metrice}
\end{align}
for $\lambda\gtrless0$ respectively. We will employ the complexification trick first introduced by \cite{NJ65} to use complex transformations in order to obtain real new metrics.

Let us discuss $\lambda<0$ and remember that we had an upper limit of $\bar r<1$ in this case.
It is convenient to use the transformation 
\ba
\bar r^2&=\cosh^2 x \left [ 1-c(y)^2\tanh^2x \right ] 
\nonumber\\
\tanh\bar t&=c(y) \tanh x
\label{itrafo}
\end{align}
leading to 
\ba
{|\lambda|\over 3}ds^2&=d x^2\!-\!{ \sinh^2 x c'(y)^2\over c^2(y)-1}\left \{
d y^2
-{C\over 3}{[c(y)^2-1][c(y)^2-{\rm coth}^2(x)]\over  c'(y)^2} d\Omega^2
\right \}
\nonumber\\
&\approx
d x^2\!-\!{ \sinh^2 x c'(y)^2\over c^2(y)-1}\left [
d y^2-{C\over 3}{[c(y)^2-1]^2\over c'(y)^2} d\Omega^2
\right ]
\label{xy}
\end{align}
where the last line is approaching for large $x$. This form (\ref{xy}) gives a variety of possible representations. Please note that with (\ref{itrafo}) we give the inverse formulas of the Mitra paradox \cite{Mi12,Mit15,Grn16} dating back to the Florides solution \cite{Fl85} how to transform FLRW metric into a Schwarzschild metric. Here we discuss a similar problem within our new solution, i.e. in which ranges an expanding or contracting universe can look static and vice versa. 

First we try a further complex transformation in (\ref{xy})  
\be
c(y)=i{\rm tan} \left (c+ \sqrt{C\over 3} \,{\rm ln}\, y\right ),\quad y=\sqrt{|\lambda|\over 3} r
\label{trafoc}
\ee 
with an arbitrary constant $c$ and $x=\sqrt{|\lambda|/3}\,t$ to bring it to the familiar form 
\be
ds^2=d t^2-a(r, t)\left (d r^2+r^2 d\Omega^2\right )
\label{ds}
\ee
with
\be
a(r, t)={C \sinh^2{\sqrt{|\lambda|\over 3}t}\over |\lambda| r^2\cos^2{(2 c+ \sqrt{C\over 3} {\rm ln} \,\sqrt{|\lambda|\over 3}r)}}.
\label{Ryx}
\ee
We see that we obtain a FLRW metric with $k=0$ like the Einstein-De Sitter universe except that the scaling radius is not only time-($t$) dependent but also space-($r$) dependent. 
We can shift the spatial dependence of the scale factor in a more standard form if we use instead of (\ref{trafoc}) the form 
\be
c(y)=i\,{\rm sinh} y,\quad y=\sqrt{|\lambda|\over 3} r
\label{trafoc1}
\ee 
to obtain 
\be
ds^2=d t^2-a(t)\left (d r^2+{C\over |\lambda|} {\rm cosh}^2\sqrt{|\lambda|\over 3} r \,d\Omega^2\right )
\label{dss1}
\ee
with
\be
a(t)= \sinh^2{\sqrt{|\lambda|\over 3}t}
\label{Ryxs1}
\ee
which allows the standard cosmological discussion with Hubble constant etc.
Please note that we have used here complex spatial coordinates in order to achieve this standard notation.

The range of allowed transformations (\ref{itrafo}) is easily seen to be $\cosh x^2=\bar r^2/(1-\tanh^2\bar t)>1$ and plotted in figure~\ref{region}. Below the upper limit $\bar r<1$ and large time $\bar t$ the range of possible transformations for time-like FLRW metrics is visible.

\begin{figure}
\centerline{\includegraphics[width=7cm]{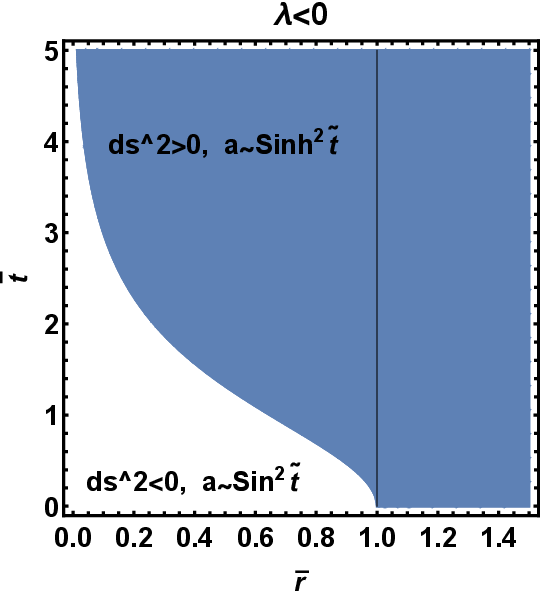}}
\caption{\label{region} The $\bar r-\bar t$-region of transformation (\ref{itrafo}) where the case $\lambda<0$ assumes a time-like FLRW metric (dark) and the region where (\ref{itrafo1}) transforms to a space-like metric (white). The upper limit of this metric was $\hat r^2=C/|\lambda|\, \bar r^2< C/|\lambda|$ visualized by the perpendicular line.}  
\end{figure}

In the missing range for smaller times we can use $\tilde x=i x=\sqrt{|\lambda|/3}\,\tilde t$ in the transformation (\ref{itrafo}) and (\ref{trafoc}) to render it real
\ba
\bar r^2&=\cos^2 \tilde x \left [ 1-c(y)^2\tan^2\tilde x \right ] 
\nonumber\\
\tanh\bar t&=c(y) \tan \tilde x
\nonumber\\
c(y)&={\rm tan} \left (c+ \sqrt{C\over 3} \,{\rm ln}\, y \right ),\quad y=\sqrt{\lambda\over 3} r
\label{itrafo1}
\end{align}
and to obtain just the negative of (\ref{ds}) with
\be
a(r, \tilde t)={C \sin^2{\sqrt{|\lambda|\over 3}\tilde t}\over |\lambda| r^2\cos^2{\left (2 c+ \sqrt{C\over 3} {\rm ln} \,\sqrt{|\lambda|\over 3}r\right )}}.
\label{Ryx1}
\ee
The negativity of the whole line element would correspond to imaginary line elements $d\tilde s^2=-d\tilde s^2$ and a space-like metric.

For the case $\lambda>0$ the transformation
\ba
\bar r^2&=\cosh^2 x \left [ c(y)^2\tanh^2x -1\right ] 
\nonumber\\
\coth\bar t&=c(y) \tanh x
\label{itrafo2}
\end{align}
brings the metric to the form
\ba
{|\lambda|\over 3}ds^2&=-d x^2\!-\!{ \sinh^2 x c'(y)^2\over 1-c^2(y)}\left \{
d y^2
-{C\over 3}{[c(y)^2-1][c(y)^2-{\rm coth}^2(x)]\over c'(y)^2} d\Omega^2
\right \}
\nonumber\\
&\approx
-d x^2\!-\!{ \sinh^2 x c'(y)^2\over 1\!-\!c^2(y)}\left [
d y^2\!-\!{C\over 3}{(c(y)^2\!-\!1)^2\over c'(y)^2} d\Omega^2
\right ]
\label{xy1}
\end{align}
where the last line is again approaching for large $x$. The further transformation (\ref{trafoc}) and $x=\sqrt{|\lambda|/3}\,t$ yields
\be
-ds^2=d t^2-a(r, t)\left (d r^2+r^2 d\Omega^2\right )
\label{ds1}
\ee
with (\ref{Ryx}). The range of allowed transformations (\ref{itrafo2}) is $\cosh x^2=\bar r^2/(\coth^2\bar t-1)>1$ and plotted in figure~\ref{region1}. We see that the space-like FLRW metric appears in this range as we had in the range of the case of $\lambda<0$. The other range can be described again by the change $\tilde x=i x=\sqrt{|\lambda|/3}\,\tilde t$ which leads to the real transformation
\ba
&\bar r^2=\cos^2 \tilde x \left [c(y)^2\tan^2\tilde x -1\right ] 
\nonumber\\&
\coth\bar t=c(y) \tan \tilde x
\nonumber\\&
c(y)={\rm tan} \left (c+ \sqrt{C\over 3} \,{\rm ln}\, y\right ),\quad y=\sqrt{|\lambda|\over 3} r
\label{itrafo3}
\end{align}
which provides the time-like FLRW metric (\ref{ds}) with (\ref{Ryx1}). The simpler choice (\ref{trafoc1}) and additionally imaginary times $x=i t\sqrt{|\lambda|/3}$ leads again to (\ref{dss1}) and (\ref{Ryxs1}) where $sinh$ has to be replaced by $sin$.

\begin{figure}
\centerline{\includegraphics[width=7cm]{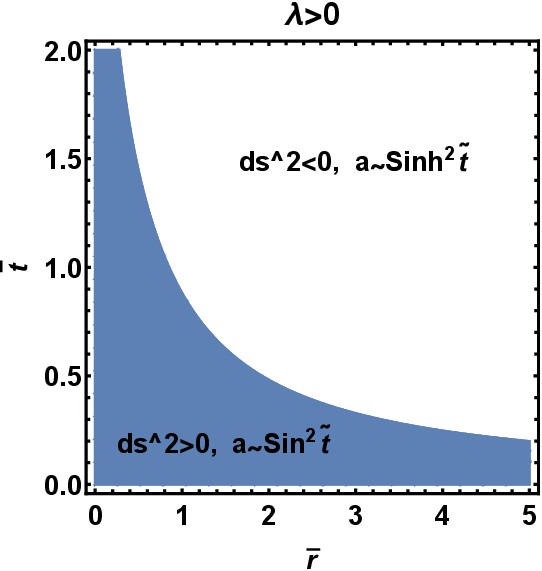}}
\caption{\label{region1} The $\bar r-\bar t$-region of transformation (\ref{itrafo2}) where the case $\lambda>0$ assumes a space-like FLRW metric (white) and the region where (\ref{itrafo1}) transforms to a time-like metric (dark).}  
\end{figure}
 
We see that for both cases $\lambda\gtrless 0$ ranges of transformations are possible to find a FLRW metric where the Einstein constant $\lambda$ takes over the role of the standard $k$ factor in Friedman cosmoses. The scaling factor $a$ becomes time and space dependent. Using complex coordinates the standard only time-dependent scaling factor can be achieved. In one range it shows oscillating behaviour and in the complimentary range an exponential increasing behaviour in time. 

\subsubsection{Morris and Thorne form} 

With the help of the transformation
\be
R(r)=-{C\over \lambda} \cos^{-2}\left [\sqrt{4 C\over 3}{\rm arcsin}\sqrt{r\over b}\right ]
\ee
we obtain from (\ref{metricd}) the metric
\be
ds^2={\rm e}^{2\Phi(r)}d t^2-a(r)\left ( {dr^2\over 1+{1\over r}}+r^2d\Omega^2\right )
\label{form}
\ee
with $a(r)=R(r)/r^2$ and $\Phi(r)=\ln\sqrt{R(r)}$ which shows that the space part can be written in a form known from Schwarzschild solution but with a space-dependent prefactor leading to the unusual features described in this paper. The form (\ref{form}) without space-dependent prefactor was used in \cite{MZH17} in order to discuss wormhole solutions of Einstein-Cartan gravity which we have already presented in the last chapter.

\subsubsection{Interchange of time and space coordinates}

Having exercised complex transformations it is revealing to consider all cases of $\lambda$ together by the complex coordinate transformation (\ref{metricd}) 
\be
\bar r&=&\sqrt{a(i\tilde t)}=\left \{\begin{array}{ll}
\sqrt{C\over |\lambda|}\sin{\sqrt{|\lambda|\over 3}i\tilde t};& \lambda<0\cr
\sqrt{C\over 3} i\tilde t;& \lambda=0\cr
\sqrt{C\over |\lambda|}\sinh{\sqrt{|\lambda|\over 3}i\tilde t};& \lambda>0\cr
\end{array}\right .
\nonumber\\
\bar t&=&i \ln \tilde r
\label{trafo}
\ee
providing the form of the metric
\ba
d s^2=d\tilde t^2-{a(i\tilde t)\over \tilde r^2} [d \tilde r^2+\tilde r^2 d\Omega^2].
\end{align}
This looks like the standard FLRW metric except that $a(i\tilde t)<0$ which means we obtain an Euclidean metric in this way. The complex old radial coordinate is now the time and the complex old time the new radial coordinate.

\subsection{Autoparallel motion}
Next we solve the autoparallel trajectories in Riemann-Cartan spacetime
\be
{\ddot x^\mu}=-\Gamma^\mu_{\nu\kappa}\dot x^\nu\dot x^\kappa.
\ee
Dots are derivatives to some proper time $\tau$ for the metric (\ref{metricd}) with $x=(t,R,\theta, \phi)$. These equations are different from the geodesic equation \cite{Ac15} which is determined by purely Riemann space-time and the neglect of contortion in (\ref{GC}). We assume a test object which can couple to torsion. The equations of motion read explicitly
\ba
\ddot t&=-\Gamma^0_{\nu\kappa}\dot x^\nu\dot x^\kappa=-\ch{0}{\nu \kappa}\dot x^\nu\dot x^\kappa
= {\dot R\dot \theta\over R} 
\nonumber\\
\ddot R&=-\Gamma^1_{\nu\kappa}\dot x^\nu\dot x^\kappa=-\ch{1}{\nu \kappa}\dot x^\nu\dot x^\kappa
=
\dot R^2{C+2\lambda R\over 2 R(b+\lambda R)}+\frac 2 3 R(b+\lambda R)(\dot \theta^2+\sin^2\theta \dot \phi^2-\dot t^2)
\nonumber\\
\ddot \theta&=-\Gamma^2_{\nu\kappa}\dot x^\nu\dot x^\kappa=-\ch{2}{\nu \kappa}\dot x^\nu\dot x^\kappa\!-\!2{\sqrt{1\!-\!C}\over \kappa} \sin\theta \dot t\dot \phi
=
\sin\theta\cos\theta \dot \phi^2\!-\!{\dot R\over R}\dot \theta\!-\!2{\sqrt{1\!-\!C}\over \kappa} \sin\theta \dot t\dot \phi
\nonumber\\
\ddot \phi&=-\Gamma^3_{\nu\kappa}\dot x^\nu\dot x^\kappa=-\ch{3}{\nu \kappa}\dot x^\nu\dot x^\kappa+2{\sqrt{1-C}\over \kappa \sin\theta } \dot t\dot \theta
=
-{\dot R\over R}\dot \phi-2\cot\theta \dot \phi\dot \theta+2{\sqrt{1-C}\over \kappa \sin\theta}  \dot t\dot \theta.
\end{align}
One sees that the autoparallel motion due to affine connection differs from the geodesic due to Levy-Cevita connection by terms $\sim\sqrt{1-C}/\kappa$ only in the two angular equations.

As usual we will consider a fixed azimuthal angle, but choose here $\theta_0\ne \pi/2$, since then the equation for $\theta$ is solved as 
\be
\phi(\tau)=-{2\over \kappa \cos\theta_0} \tau+\tau_0
\label{phit}
\ee
providing a linear connection between the proper time $\tau$ and the angle $\phi$.
Then the equation for $t$ is integrated to provide
\be
R\dot t=F={\rm const}
\label{Rt}
\ee
which allows to close the equation for
$\bar R=|\lambda|R/C$ as
\be
\ddot {\bar R}=-{1\pm \bar R\over \bar R}+{1\pm 2 \bar R\over 2 \bar R(1\pm \bar R)}\dot {\bar R}^2
\label{R}
\ee
for $\lambda\gtrless 0$ and 
where we have used the scaling 
\be
\bar \tau=\tau {F} |\lambda|\sqrt{{2\over  3 C}[\kappa^2+4(C-1)\tan^2\theta_0]}.
\label{121}
\ee

The equation (\ref{R}) and following (\ref{Rt}) can be solved exactly as 
\ba
\bar R&=\pm {2(z-1)\over 2\mp c_1 z}\nonumber\\
 z&=\tanh \sqrt{c_1 C\over 3} t=\sqrt{2\over c_1}\tan({\rm h})\sqrt{c_1\over 4} (\bar \tau+c_2)
\label{139}
\end{align}
with $\tanh/\tan$ for $\lambda\gtrless 0$ respectively with the two integration constants $c_1$ and $c_2$. The integration constant $c_2$ is an irrelevant shift in the parameter $\tau$ while the meaning of $c_1$ depends on the actual chosen coordinate system.

With the help of (\ref{phit}) we can give a parametric plot $R(\phi)$ in figure~\ref{rphi_lg0_cl0} for $\lambda>0$ and in figure~\ref{rphi_ls0_cl0} for $\lambda<0$. For large negative $\bar t$ or $\phi$ the trajectories for a fixed $\theta_0$ approaches circles. For large $t$ the trajectories approach the radius of $0$ except $c_1=2$ which provides a fixed radius of $\bar R=1$. The diverging radius appears for angles corresponding to $z=2/c_1$. Analogous discussion can be performed for $\lambda<0$ as seen in figure~\ref{rphi_ls0_cl0}.

\begin{figure}
\centerline{
\includegraphics[width=7cm]{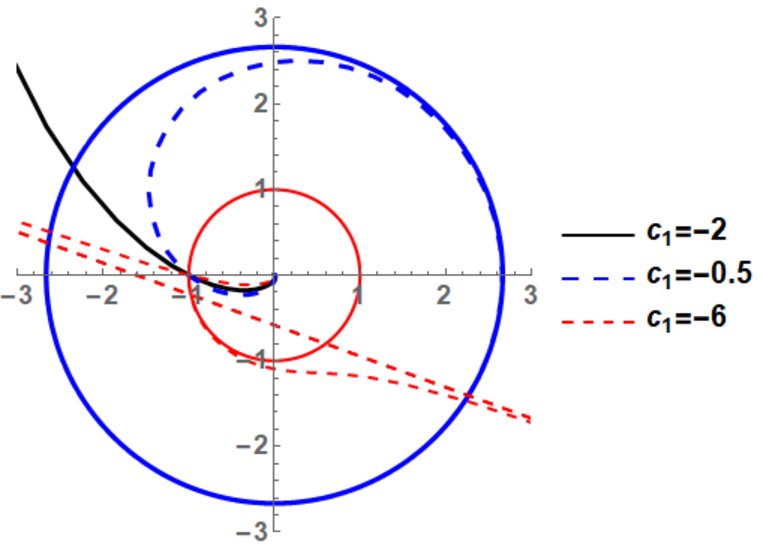}
\includegraphics[width=7cm]{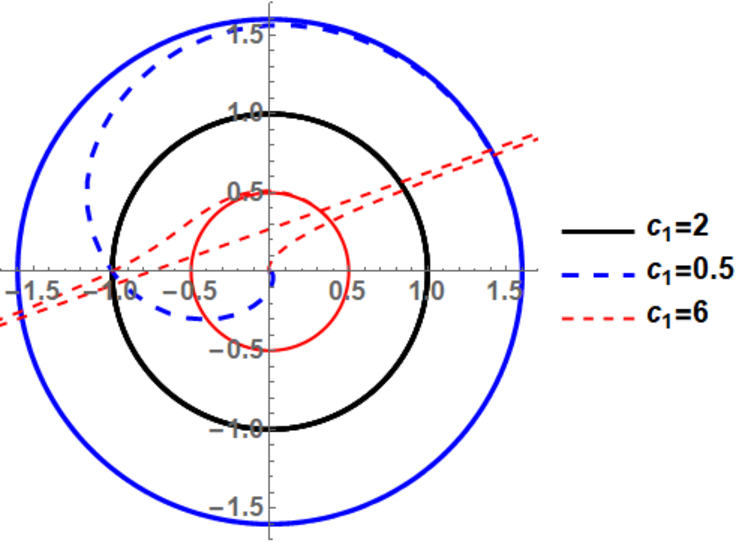}}
\caption{\label{rphi_lg0_cl0} The parametric plot $\bar R(\phi,\theta_0)$ of (\ref{139}) for $\lambda>0$ and different values of the integration constant $c_1$. In case $c_1=-2$ an infinite trajectory appears. For angles $z=2/c_1$ the trajectory diverges as well.}  
\end{figure}

\begin{figure}
\centerline{\includegraphics[width=7cm]{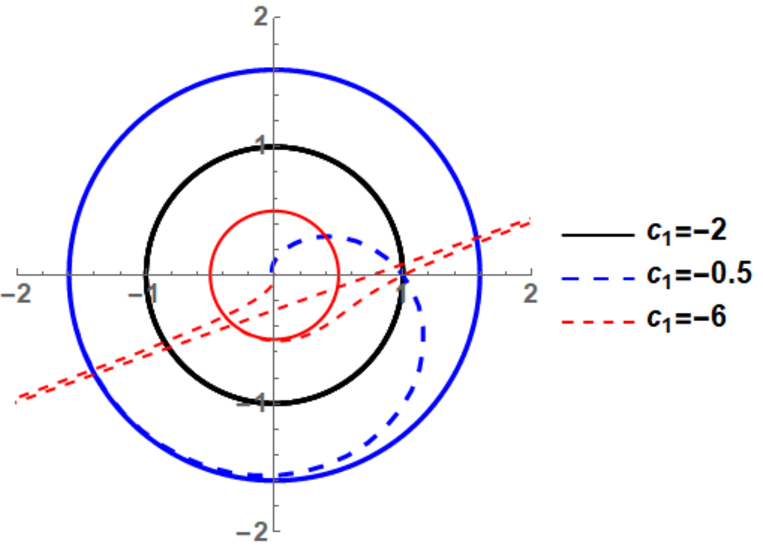}
\includegraphics[width=7cm]{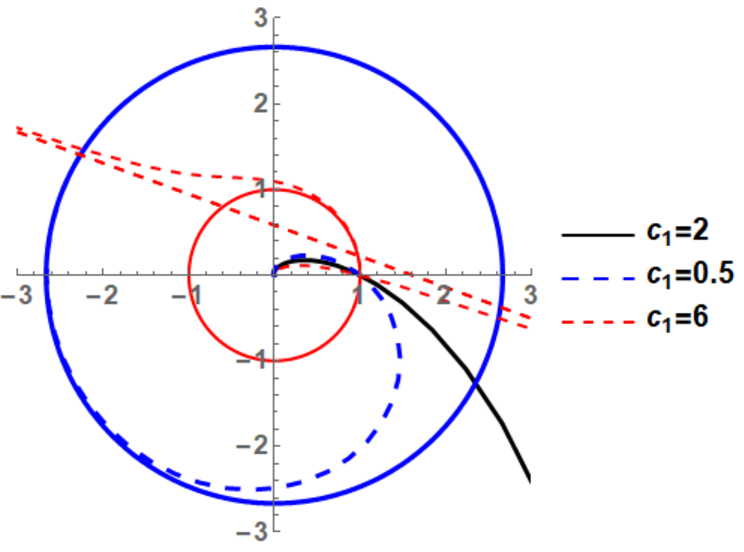}}
\caption{\label{rphi_ls0_cl0} The parametric plot $R(\phi,\theta_0)$ of (\ref{139}) for $\lambda<0$ and different values of the integration constant $c_1$. For $c_1=2$ an infinite trajectory appears as well as for angles $z=-2/c_1$.}  
\end{figure}

Let us discuss some possibilities for the constant $c_1$. It is easily seen that
\be
\dot R^2=(\pm1+R)(2+c_1 R)
\ee
holds and for $R=R[r(t)]$ we can interpret
\be
{\dot r^2\over 2}={\dot R^2\over 2 R'^2}={(\pm 1+R)(2+c_1 R)\over 2 R'^2}=-V^{\rm eff}(r)
\ee
as effective gravitational potential. Choosing $R=\sinh^2(r/2^{3/2})$ for $\lambda>0$ and $R=\sin^2(r/2^{3/2})$ for $\lambda<0$ we obtain
\be
{\dot r^2\over 2}\mp {2\over \sin(h)^2{r\over 2^{3/2}}}=\pm c_1
\ee
which one can verify as first integral of (\ref{R}) also directly by substitution. In this coordinate choice $\pm c_1$ plays the role of total energy. Among the many possibilities we want to highlight some special choices of coordinates
\ba
&R(r)=1-{a\over r}:
&
V^{\rm eff}&=&&
\left \{\!\begin{array}{ll}{-r^2\over 2 a^2}(2 r\!-\!a )[(2\!+\!c_1) r-c_1 a]&\lambda\!>\!0\cr
-{r^2\over 2 a^2}[c_1 a\!-\!r(2\!+\!c_1)]&\lambda\!<\!0
\end{array}\right .
\label{pot0}
\\
&R(r)=\frac 2 3 r^{3/2}:
&V^{\rm eff}&=&&-\frac 1 r -\sqrt{r}{2\pm c_1\over 3}-{2 c_1\over 9}r^2,\quad \lambda\gtrless 0
\label{pot1}
\\
&R(r)=\!\pm {\rm e}^{\frac 2 3 r^{3/2}}, \, \lambda\gtrless 0:
&V^{\rm eff}&=&&-{\left (1\!+\!{\rm e}^{-\frac 2 3 r^{3/2}}\right )\left (c_1 \!\pm\! {2}{\rm e}^{-\frac 2 3 r^{3/2}}\right ) \over 2 r}
\nonumber\\
&&&\approx&& {\mp2\!-\!c_1\over r}\!+\!\left (\pm 2\!+\!{c_1\over 3}\right)\sqrt{r}\!+\!o(r^2).
\label{pot}
\end{align}
One sees from (\ref{pot0}) that the simple demand of obtaining gravitational force as we did in (\ref{53})  would not lead to an effective gravitational potential in geodesic motion. The choice of (\ref{pot1}) yields a gravitational potential but with a confining large distance behaviour. Only the last choice (\ref{pot}) vanishes at large distance and is illustrated in figure \ref{Veff_lg0}. Please note that we have scaled out the "angular momentum" equivalent $F$ of (\ref{Rt}). Dependent on the parameter $c_1$ we see that the gravitational potential gets a maximum where inflection is possible. This is actually the case for $\lambda>0$ when we fix as special choice $c_1=-1$ in order to reproduce the gravitational potential at short distances. The case of $\lambda<0$ is almost inverse to the one $\lambda>0$ but the special choice $c_1=3$ in order to reproduce the gravitational force at short distances does not yield any maximum.

\begin{figure}[h]
\begin{center}
\includegraphics[width=8cm]{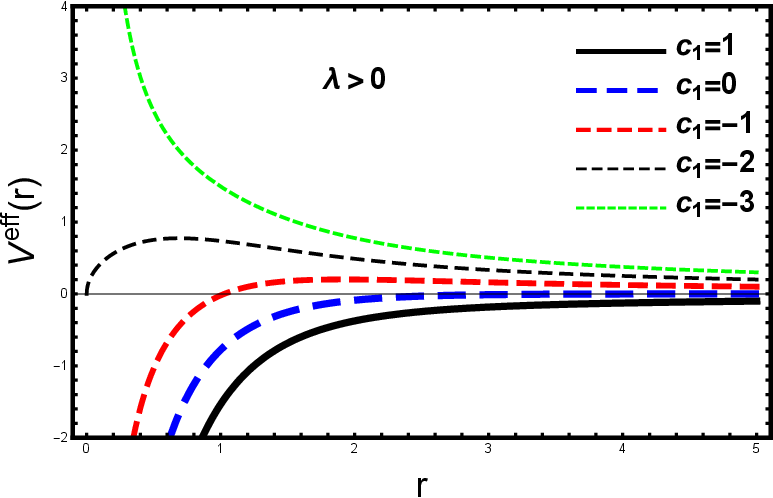}\\
\includegraphics[width=8cm]{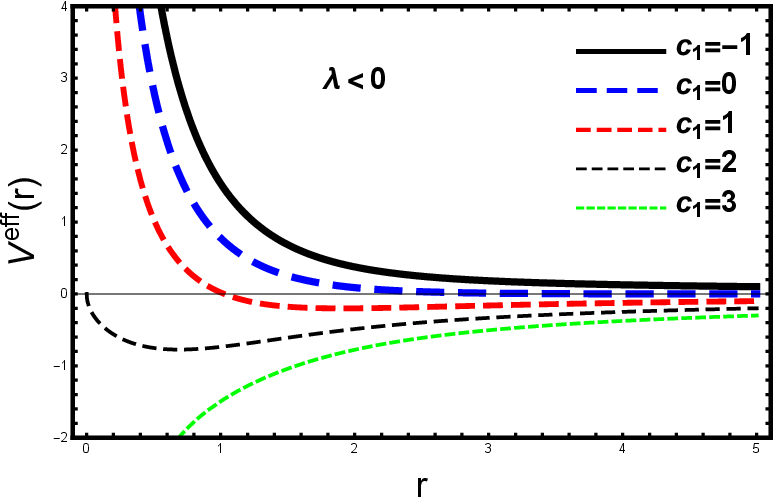}\\
\includegraphics[width=8cm]{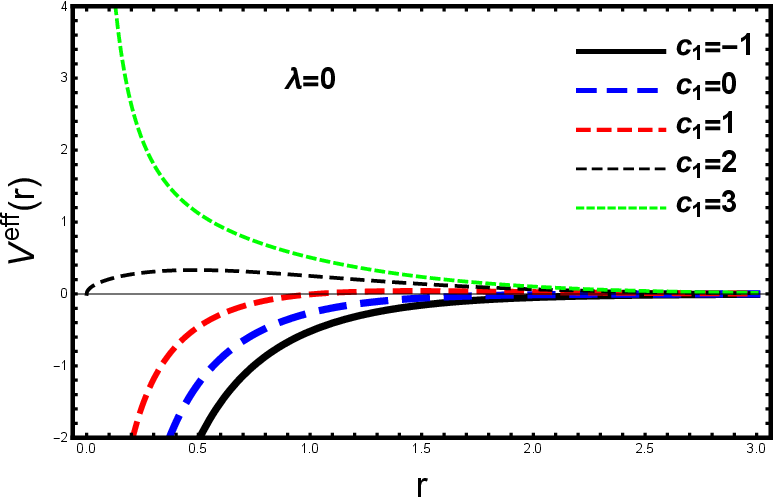}
\end{center}
\caption{\label{Veff_lg0} The effective gravitational potential (\ref{pot}) for $\lambda>0$ (above) and $\lambda<0$ (middle) and (\ref{pot2}) for $\lambda=0$ (below) for different values of the integration constant $c_1$.}  
\end{figure}

For completeness let us present shortly the results for $\lambda=0$. For the scaled coordinate $R=C\bar R$ and proper time, to set $\lambda=1$ in (\ref{121}), we get the differential equation  
\be
\ddot {\bar R}=-{1\pm \bar R\over \bar R}+{\dot {\bar R}^2\over 2 \bar R}
\ee
which has a first integral
\be
{\dot {\bar R}^2\over \bar R}-{2\over \bar R}=c_1.
\ee
The coordinate choice $R=r^2/8$ provides 
\be
{\dot r^2\over 2}-{16\over r^2}=c_1
\ee
and therefore an inverse squared effective potential and the constant $c_1$ to be interpreted as energy. A asymptotically vanishing potential with $1/r$ behaviour at short distances is reached for the choice (\ref{pot}) which reads here
\be
V^{\rm eff}&=&-{{\rm e}^{-\frac 2 3 r^{3/2}}\left (c_1 \!+\! {2}{\rm e}^{-\frac 2 3 r^{3/2}}\right ) \over 2 r}
\approx
{c_1-2\over 2 r}+\frac 1 3 \left (4-{c_1}\right)\sqrt{r}+o(r^2)
\label{pot2}
\ee
presented in figure~\ref{Veff_lg0} as well. The special choice $c_1=0$ does not produce a second maximum.

The here presented analysis considers explicitly the presence of a cosmological constant. Other dynamical models with curved space time and torsion in dependence on the cosmological constant can be found in \cite{Br08}.

\subsection{Weyl tensor}
It is customary to calculate the traceless Weyl tensor 
\ba
W_{abcd}&=&P_{abcd}-\frac 1 2 \left (g_{ac}P_{bc}+g_{ad}P_{ac}-g_{ad} P_{bc}-g_{bc} P_{ad}\right )
+{P\over 6}\left (g_{ac}g_{ad}+g_{ad}g_{bc}\right )
\end{align}
in order to classify the solution \cite{Ah19,StK03,GP09}. Using the Newman-Penrose formalism \cite{Pi57,NP62} we calculate for the new second solution (\ref{m1}) with (\ref{sol3}) the nulltretrads \cite{Ah19} 
\ba
\eta_1&=l={1\over \sqrt{2}}(\omega_0-\omega_1),
&
\, \eta_2
&
=n={1\over \sqrt{2}}(\omega_0+\omega_1),
\,
\nonumber\\ 
\eta_3
&
=m={1\over \sqrt{2}}(\omega_2+I \omega_3),
&
\, \eta_4
&
=\bar m={1\over \sqrt{2}}(\omega_2-I \omega_3)
\end{align}
from the orthonormal tetrads
\ba
\omega_0&=(r,0,g_{02} r,0), \quad \omega_1=(0,{\sqrt{3}\over \sqrt{{C\over 1+g_{02}^2}+\lambda r^2}}, 0,0),
\nonumber\\
\quad
\omega_2
&
=(0,0,\sqrt{1+g_{02}^2}r,0),\quad \omega_3=(0,0,0,r\sin\theta).
\end{align}
The five self-dual expansion components of the Weyl tensor are
with $\tilde W_{ijkl}=W_{abcd} \eta^a(i)\eta^b(j)\eta^c(k)\eta^d(l)$
\be
\Psi_0&=&-W_{abcd}l^a m^b l^c m^d=-\tilde W_{1313}=0
\nonumber\\
\Psi_1&=&-W_{abcd}l^a n^b l^c m^d=-\tilde W_{1213}=0
\nonumber\\
\Psi_2&=&-W_{abcd}l^a m^b \bar m^c n^d=-\tilde W_{1342}=-{1\over 6(1+g_{02}^2) r^2}
\nonumber\\
\Psi_3&=&-W_{abcd}l^a n^b \bar m^c n^d=-\tilde W_{1242}=0
\nonumber\\
\Psi_4&=&-W_{abcd}n^a \bar m^b n^c \bar m^d=-\tilde W_{2424}=0
\ee
and provides the Petrov-D classification. This is the same class as the different solution found by \cite{Ka19}. This solution means that two double principal null directions appear like in the gravitational fields of massive objects.

One might argue now that the Weyl tensor and the following classification scheme is valid only for the Riemann tensor and not for the curvature tensor including torsion. One can repeat the analysis calculating the Weyl tensor from the curvature tensor which means to replace the Christoffel symbols by the affine connection $\Gamma$ according to $(\ref{GC})$. The result is
\be
\Psi_0&=&\Psi_1=\Psi_3=\Psi_4=0
,\,
\Psi_2
=
-{2(1+C)+\kappa^2\over 6(1+g_{02}^2)\kappa^2 r^2}
\ee
leading to the same Petrov-D solution. 

\section{Summary}

The Einstein-Cartan equations resulting from lowest-order action are considered. An explicit restriction on the spin tensor is derived in order to keep the consistence of equations and conservation laws. Inside matter the standard gravitational equations are obtained with torsion including the generalized Oppenheimer-Volkov equation. Outside matter the equations are solved exactly. Besides the spherically symmetric Schwarzschild solution a second non-spherical one appears which is proven to be unique by testing various non-spherical parametrizations of the metric. The found second solution is characterized by an equivalent spin component which is proportional to the inverse squared of the Einstein coupling constant and appears not as an independent variable. Instead it is completely determined by the Einstein-Cartan equations together with the metric. It does not follow as an additional parametrically spin added to the gravitation but is a consequence of consistence of equations. 

The new solution possesses unusual properties. The comparison with other exact solutions show that it is not a part of known ones. Special transformations are constructed which allows to specify areas of time- and space-like FLRW forms but with space-dependent form factors. The usual forms of only time-dependent scaling is only achieved by complex coordinate transformations which rises the question of interpretation. We see it as sign of a disjunct solution from known Friedman cases. The cosmological constant takes the role of the standard $k$ parameter of Friedman cosmoses. Depending on the sign of the cosmological constant an open or closed cosmos is obtained for this new metric. General transformation formulas are presented which allow to transform regions of static Schwarzschild solutions into time-dependent FRLW metrics and vice-versa. The possibilities of wormholes are discussed and the effects of unusual curvature in this solution is shown like a permanent curvature also at small distances. 

The autoparallel equations are solved exactly and are compared with geodesic motion in this new solution dependent on the sign of the cosmological constant. Appropriate coordinate systems are constructed such that gravitational forces appear at short distances and the consequent effective gravitational potentials are derived. They show a deflection behaviour dependent on the total energy in the system. 

Finally the Weyl tensor is calculated and the presented exact solution is identified as Petrov-D type like it appears in the neighborhood of massive objects.

\section*{Acknowledgements}
The author likes to thank Emmanuel Vignes for pointing out some misprints which have been published as corrigendum in Class. Quant. Grav. 40 (2023) 029501. Further misprints and wrong signes have been found by Daniel Janda within his thesis work supervised by Filip Petr\'asek for which findings the author likes to express his gratitude.

\section*{References}
\bibliography{entropy,bose,kmsr,kmsr1,kmsr2,kmsr3,kmsr4,kmsr5,kmsr6,kmsr7,delay2,spin,spin1,refer,delay3,gdr,chaos,sem3,sem1,sem2,short,cauchy,genn,paradox,deform,shuttling,blase,spinhall,spincurrent,tdgl,pattern,zitter,graphene,quench,msc_nodouble,iso,march,weyl,anomal,darkmatter,rel}

\bibliographystyle{unsrt}

\end{document}